\documentclass[aps,prb,reprint,showpacs]{revtex4-1}

\usepackage{bm,color,amsmath,amssymb,mathrsfs,latexsym,graphicx,psfrag}
\usepackage[small]{subfigure}

\def\be{\begin{equation}}
\def\ee{\end{equation}}
\def\ba{\begin{eqnarray}}
\def\ea{\end{eqnarray}}

\begin{document}

\title{Band-structure effects on superconductivity in Hubbard models}
\author{Weejee Cho${}^1$}
\author{Ronny Thomale${}^{2,3}$}
\author{Srinivas Raghu${}^1$}
\author{Steven A. Kivelson${}^1$}

\affiliation{${}^1$ Department of Physics, Stanford University, Stanford, CA 94305}
\affiliation{${}^2$ Institut de th\'eorie des ph\'enom\`enes physiques, \'Ecole Polytechnique F\'ed\'erale de Lausanne (EPFL), CH-1015 Lausanne}
\affiliation{${}^3$Institute for Theoretical Physics and Astrophysics, University of W\"urzburg, D 97074 W\"urzburg}

\date{\today}
\begin{abstract}
We study the influence of the band structure on the symmetry and superconducting transition temperature in the (solvable) weak-coupling limit of the repulsive Hubbard model. Among other results we find that (1) as a function of increasing nematicity, starting from the square-lattice (zero nematicity) limit where a nodal $d$-wave state is strongly preferred, there is a smooth evolution to the quasi-1D limit, where a striking near-degeneracy is found between a $p$-wave- and a $d$-wave-type paired states with accidental nodes on the quasi-one-dimensional  Fermi surfaces---a situation that may be relevant to the Bechgaard salts. (2) In a bilayer system, we find a phase transition as a function of increasing bilayer coupling from a $d$-wave to an $s_\pm$-wave state reminiscent of the iron-based superconductors. (3)  When an antinodal gap is produced by charge-density-wave order, not only is the pairing scale reduced, but the symmetry of the pairs switches from $d_{x^2-y^2}$ to $d_{xy}$; in the context of the cuprates,  this suggests that were the pseudo-gap entirely due to a competing CDW order, this would likely cause a corresponding symmetry change of the superconducting order (which is not seen in experiment).
\end{abstract}
\maketitle

\section{Introduction}
\label{sec:intro}
Electronic pairing mechanisms have been extensively studied in the context of high-temperature superconductivity, with much focus on the prototypical Hubbard model with a repulsive on-site interaction. \cite{[{For a review of the strong-coupling perspective, see }][{.}]lee-06rmp17,[{For a more general review, see }][{.}]scalapino12rmp1383} It is challenging to obtain unequivocal results for this model owing to the fact that there exist no well-controlled solutions (in more than one dimension) in the physically most relevant regime, where the strength of the interaction is comparable to the bandwidth. We can, however, use the perturbative renormalization group (RG) method to obtain asymptotically exact results\cite{raghu-10prb224505} in the weak-coupling (small $U$) limit, in which unconventional superconductivity  occurs without any fine tuning. One might hope that insights from the weak-coupling limit, such as  trends in $T_c$, and the structure of the order parameter,  may carry over qualitatively to real materials.  This notion is supported by the fact that a variety of different physically motivated approximate calculations based on weak-coupling reasoning  seem to give consistent results when extrapolated to intermediate couplings \cite{raghu-10prb224505,scalapino12rmp1383} (see Sec. \ref{discussion}).

That unconventional superconductivity can arise from a purely repulsive interaction has been known for a long time. \cite{kohn-65prl524,berk-66prl433,emery-86prb7716,scalapino-86prb8190,kagan-88jetplett614,baranov-92ijmp2471} Generically, in the weak-coupling limit, the bare interaction is purely repulsive while second-order processes, although typically attractive, are much weaker. However, when the bare interactions are short-ranged, the induced attraction can overscreen the bare repulsion. For the case of the Hubbard model, the bare interaction operates only on-site; the second-order induced attraction is nonlocal and therefore operates in non-trivial channels where the bare interaction has no effect. The effective interaction inherits its $\textbf{k}$-space structure from virtual particle-hole excitations, whose properties are in turn sensitive to the band structure, not only at the Fermi energy but further away from it as well.

Specifically, we consider a reference ``undistorted'' Hubbard model defined on a square lattice, and then study the effect of various distortions of the band structure on the superconducting $T_c$. In the context of the cuprates, these distortions can be thought of as arising from the presence of specific forms of ``competing''  orders, including various density wave, orbital current, and electron nematic phases that may play a role in the pseudogap regime. \cite{chakravarty-01prb094503,kivelson-98nature550,varma06prb155113,dicastro-96prb16126} To address the formation of these phases, and their relationship to superconductivity, one necessarily must solve the intermediate-coupling problem which has no small parameters.  However, deep inside such phases, these non-superconducting orders can be represented as static mean-fields that reconstruct the bare band structure.

Our results provide a general perspective on the mechanism of unconventional superconductivity.  In the limit of a strong nematic distortion, the band structure is that of a quasi-1D superconductor; as a function of increasing strength of the coupling between the two planes in a square lattice bilayer, we find a transition from $d$- to $s_\pm$-wave pairing symmetry.
  {\it The fact that  we can follow the evolution of the superconducting state from the undistorted limit in which it has  simple  $d$-wave pairing, of the sort found in the cuprates, to an s$_\pm$ state, of the sort thought to occur in the Fe-based superconductors, or to a quasi-1D case in which there is a near degeneracy between a singlet and a triplet paired state with ``accidental nodes,'' which may be relevant to the Bechgaard salts or, possibly even to Sr$_2$RuO$_4$, suggests a unified understanding of the origin of unconventional pairing across a broad range of materials. }
Although this mechanism  is loosely  related to the ``spin fluctuation exchange'' that has been widely discussed, \cite{scalapino12rmp1383,berk-66prl433,emery-86prb7716,scalapino-86prb8190,varma-86prb6554} in the limit studied here there is no well-defined collective mode that can be thought of as ``the glue;'' \cite{anderson22062007} rather,  the pairing is a result of overscreening by the entire band.

This paper is organized as follows: In Sec. \ref{results}, we summarize the key results obtained from our study. In Sec. \ref{methods}, we review the perturbative renormalization group (RG) method developed in Ref.~\onlinecite{raghu-10prb224505}, and present a generalization for multiband cases. In addition, we review the general features of gap structures from unconventional pairing, and establish the physical considerations that determine the preferred superconducting order parameter symmetry for a given Fermiology. In Sec. \ref{bandstructures}, we discuss various model problems of band-structure effects on superconductivity. In Sec. \ref{discussion}, we speculate more broadly on the implications of our results for real materials, where interactions are never weak, so we are forced, without formal justification, to extrapolate  to intermediate couplings.

\section{Results}
\label{results}

\subsection{Optimal band structure for superconductivity}

There is considerable ambiguity in how to define the optimal condition for superconductivity, even in principle, since the answer depends on what is held fixed.  This issue arises even more clearly when real materials or composites are considered, where many microscopic interactions are changed when any macroscopic characteristic is varied.  Thus, we have focused on a number of {\it qualitative} features that emerge from an exploration of $T_c$ as a function of specific parameters that alter the band structure in the various models we have studied.

{\it \bf Role of Van Hove points: }
  The square lattice affects the band structure near half-filling in a way that is highly conducive to $d_{x^2-y^2}$-wave superconductivity. As has been elucidated in many places, \cite{dagotto-94rmp763,baranov-92ijmp2471,kagan-88jetplett614,zanchi-00prb13609,halboth-00prb7364,honerkamp-01prb035109} this is a combined effect of a maximum in the particle-hole susceptibility near $(\pi,\pi)$ and the fact that this same vector connects the ``antinodal'' portions of the Fermi surface that pass near the Van Hove points, i.e., ${\bf X}=(\pi,0)$ and $\bar{\bf X}=(0,\pi)$ where the Fermi velocity vanishes, leading to a logarithmically divergent density of states when the chemical potential is tuned to them.

However, as can be seen in Figs. \ref{nematicpairing} and \ref{bilayerpairing}, the dimensionless pairing interaction is substantial for a relatively broad range of chemical potentials in the neighborhood of the critical value. This reflects the fact that the Van Hove points are ``points'' on the Fermi surface whereas the condensation energy involves the entire Fermi surface.  The precise significance of the Van Hove point would be  further reduced by quasiparticle lifetime effects, either from higher order processes in powers of $U/t$ or  due to the presence of weak disorder.

{\it \bf  Multiple Fermi pockets:}
Multiple Fermi surfaces occur in many materials as a multi-orbital effect, but they can just as easily appear in the Hubbard model with
multiple sites in the unit cell, which is often a consequence of translation symmetry breaking orders. In particular, as is illustrated in the bilayer example (see Figs. \ref{bilayerpairing} and \ref{bilayerFS}) when electron-type and hole-type pockets are present it can be energetically preferable for the pairing gap to change sign between these pockets, but to be nearly constant (nodeless) on each pocket.  Somewhat less intuitive, however, is the diverse behaviors that are possible even as the size of a Fermi pocket tends to zero:  As shown for the case of the $(\pi,\pi)$-CDW (discussed in Section \ref{pipiDW}), the pairing strength need not vanish and the preferred gap can have sign changes (i.e., be nodal) as a function of angle around the pocket, even in this limit. \cite{chubukov-12annurevconmatphys020911}  This follows from the fact that the density of states of an elliptical Fermi surface in 2D is independent of its enclosed area (i.e., $|p_F|^2$).

{\it\bf Optimal inhomogeneity for superconductivity:}
It has been proposed that there is an ``optimal inhomogeneity for superconductivity,'' a notion that has been investigated with contradictory results  using various approximate methods \cite{yao-07prb161104r,karakonstantakis-11prb054508,baruch-10prb134537,tremblay-11prb054545,maier-08prb020504} in the context of the checkerboard Hubbard model---defined in Section \ref{checkerboard}.  Unambiguously in the weak-coupling limit we find that the dimensionless pairing interaction, $\lambda$, is a strongly increasing function of the checkerboard potential, although for reasons that are somewhat trivial.

\subsection{Interpolating between limiting cases}

{\bf From the square symmetric to quasi-1D limit:}
Nematic order  spontaneously breaks the $C_4$ rotational symmetry of the square lattice to $C_2$;  at a band-structure level, the degree of nematicity is the difference between the hopping amplitudes in the $x$ and $y$ directions. Over the entire range of nematicity, the dominant spin-singlet superconducting instability remains a $d$-wave-type singlet with four gap nodes on the Fermi surface (see Figs. ~\ref{nematicpairing} and ~\ref{nematicFS}), which are required by symmetry only in the limit of zero nematicity. This smooth evolution highlights the unity of mechanism involved in the square lattice and 1D limits, and identifies the ``accidental'' gap nodes in the quasi-1D system as vestigial $d$-wave nodes.  For large nematicity there is also a spin triplet ($p$-wave) paired state that is nearly degenerate with the singlet state.  There is considerable experimental evidence that both these features of the superconductivity may be relevant to the Bechgaard salts. \cite{[{For a review, see }][{.}]brown-12jpsj011004,bourbonnais08,doiron-leyraud-09prb214531} Our finding of an approximate degeneracy between $d$-wave and $p$-wave states might suggest further investigations such as applying magnetic fields to break this degeneracy.

{\bf From $d$- to $s_{\pm}$-wave pairing in a bilayer model:}
In a bilayer square lattice model near half filling, as a function of interlayer tunneling, the band structure evolves from a weak-tunneling regime in which two nearly identical electron-like Fermi surfaces enclose the $\Gamma$ point, to a strong-tunneling regime with an electron and hole pocket enclosing, respectively, the $\Gamma$ and $M$ points. The latter case shares a salient feature of the fermiology of the pnictide superconductors. \cite{mazin-08prl057003} As shown in Fig. \ref{bilayerpairing} (and consistent with earlier results of Maier and Scalapino \cite{maier-11prb180513}), a phase transition occurs for intermediate tunneling where  the symmetry of the order parameter changes from $d$-wave for small bilayer coupling to $s_\pm$ for large tunneling. \cite{scalapino-92prb5577,mazin-95prl2303} However, the basic pairing ``mechanism''  remains the same illustrating the principle that different order parameter symmetries can emerge from the same underlying mechanism, depending on the band structure.

\subsection{Insufficiency of competing orders to fully explain the ``pseudo-gap'' phenomenology}
The pseudogap, \cite{norman-05ap715} which in the hole-doped cuprates dominates the ``normal'' state above the superconducting $T_c$, has a $\textbf{k}$-space structure that is of $d$-wave type;  it is vanishingly small in the ``nodal region'' (i.e.,  near the points on the Fermi surface at which the $d$-wave nodes in the quasiparticle dispersion appear in the superconducting state) and it is largest in the anti-nodal regions.   In the superconducting state below T$_c$, the gap takes on a standard $d_{x^2-y^2}$-wave form, as found for materials that are not too strongly underdoped.  These facts suggest that the pseudo-gap is somehow intimately related to the $d$-wave superconducting gap.
Conversely, the pseudo-gap has a different temperature and doping dependence than the near-nodal superconducting gap, and shares other features suggestive of a distinct, non-superconducting order.

Given that many different types of order appear to be in close competition in the cuprates, so that the various orders are more ``intertwined'' \cite{berg-09njp115004} than simply ``competing," it may not be possible to unambiguously define the extent to which  the pseudo-gap reflects precursor pairing correlations versus the effect of an actual or incipient density-wave order. On the basis of the weak-coupling analysis, we are able to highlight a possible inconsistency with a viewpoint that attributes the pseudo-gap {\em entirely} to a nonsuperconducting order:

Suppose, that the pseudo-gap arose entirely from some form of particle-hole ordering---for example, from $(\pi,\pi)$-CDW order.  The fact that the pseudo-gap eliminates states near the $X$ points in the BZ, i.e., precisely the states that most strongly contribute to the  $d_{x^2-y^2}$ pairing, certainly leads to a suppression of $T_c$, as is clear in the model calculation shown in Fig.~\ref{dwpairing08}. It also leads to a {\it change in symmetry} of the dominant order parameter. Indeed, in the model considered, the dominant instability switches from $d_{x^2-y^2}$- to $d_{xy}$-wave superconductivity once the CDW order is strong enough to open a complete antinodal gap. No such change in symmetry has been seen in the cuprates.

\section{Model and Methodology}
\label{methods}
It is well established that the effective field theory involving weakly interacting degrees of freedom close to a Fermi surface can be treated using a perturbative renormalization group (RG) approach. To connect this to the behavior of a microscopic model, Ref.~\onlinecite{raghu-10prb224505} introduced a two-step analysis where in the first step, the low-energy effective action is derived from the Hubbard model, and in the second step, the method of Shankar and Polchinski \cite{polchinski92,shankar94rmp129} is implemented to calculate the RG flow.

In the first step, the effective action is obtained by integrating out all high-energy modes outside an asymptotically thin energy shell of width $\Omega_{0}$ around the Fermi energy using perturbation theory. There is an arbitrariness in the choice of $\Omega_{0}$. For a system with bandwidth $W$ and density of states at the Fermi surface, $\rho$, $\Omega_0$ must be taken large enough,  $\Omega_{0}\gg W e^{-1/\rho U}$, to avoid the breakdown of perturbation theory which would otherwise arise due to logarithmically divergent terms in the series. At the same time, $\Omega_0$ must be small enough, $\Omega_{0} \ll \rho U^{2}\ll W$ that all quantities with non-singular dependences on $\Omega_{0} $ can be replaced by their values evaluated at $\Omega_0=0$ and, moreover, errors associated with linearizing the dispersion relation near the Fermi energy can be neglected.

In the next step, the RG flow is calculated from the effective action. The superconducting transition temperature is identified with the energy scale where the dimensionless coupling constant in the Cooper channel grows to order unity. It can be shown that the final result is independent of $\Omega_{0}$. This relatively simple analysis is justified because in the weak-coupling limit, superconductivity is the only generic instability of the Fermi liquid, as long as we avoid perfect nesting in the particle-hole channel or Van Hove singularities.  Although our method breaks down exactly at these singularities, it is still valid as long as we stay away from them by more than $O(e^{-1/\rho U})$. When we discuss the ``behavior at a singularity,'' it should be kept in mind that there always exists a parametrically narrow region around it, which is not included in our description.

\subsection{Model}
To go through the details of the procedure described above, we consider the most general Hubbard Hamiltonian encompassing all cases studied in this paper. Its quadratic part is a tight-binding model with one orbital per atom and one or more atoms per unit cell. There is a local Hubbard interaction (which for simplicity we take to be the same on all sites, even if there is more than one per unit cell) which gives an energy penalty of amount $U$ whenever two electrons reside on the same atom. In momentum space, the Hamiltonian reads:
\begin{equation}
\begin{split}
H &= H_{0} + V \, ,\\
H_{0} &= \sum_{\textbf{k},\tau,\tau^{\prime},\sigma} t_{\tau\tau^{\prime}}^{( \sigma)}(\textbf{k})\, c_{\textbf{k}\tau\sigma}^{\dagger}c_{\textbf{k}\tau^{\prime}\sigma} \, ,\\
V &= \frac{U}{N}\sum_{\{\textbf{k}_{i}\},\tau} c_{\textbf{k}_{1}\tau\uparrow}^{\dagger} c_{\textbf{k}_{2}\tau\downarrow}^{\dagger} c_{\textbf{k}_{4}\tau\downarrow} c_{\textbf{k}_{3}\tau\uparrow}\, ,
\label{model}
\end{split}
\end{equation}
where $\textbf{k}_{4}=\textbf{k}_{1}+\textbf{k}_{2}-\textbf{k}_{3}$, $N$ is the number of unit cells in the system, and $\tau$ and $\tau^{\prime}$ denote sublattices in a unit cell. No spin-flip term exists
in $H_{0}$ as (for simplicity) we have not included  effects of spin-orbit coupling, and the only magnetically ordered states we will treat have collinear spin ordering. $t_{\tau\tau^{\prime}}^{(\sigma)}(\textbf{k})$ is a Hermitian matrix which is diagonalized by the set of orthonormal eigenvectors $\alpha_{\tau}^{(\sigma)}(n,\textbf{k})$ with corresponding eigenvalues $\epsilon_{\sigma}(n,\textbf{k})$. These eigenvalues are the energy dispersion of the non-interacting band structure and $n$ is the band index. Henceforth, we will use the compressed notation $k\equiv (n,\textbf{k})$. We can rewrite the Hamiltonian as
\ba
&&H_{0} = \sum_{{k}\sigma} \epsilon_{\sigma}({k})\, c_{{k}\sigma}^{\dagger}c_{{k}\sigma} \, ,\\
&&V = \frac{U}{N}\sum_{\{{k}_{i}\}} M({k}_{1},{k}_{2};{k}_{3},{k}_{4})
 c_{{k}_{1}\uparrow}^{\dagger} c_{{k}_{2}\downarrow}^{\dagger} c_{{k}_{4}\downarrow} c_{{k}_{3}\uparrow}\, ,
\nonumber
\ea
where
\ba
&&M(k_1,k_2;k_3,k_4)\\
\label{bareint}
&&\equiv\sum_{\tau}\alpha_{\tau}^{(\uparrow)}({k}_{1})^{\ast} \alpha_{\tau}^{(\downarrow)}   ({k}_{2})^{\ast} \alpha_{\tau}^{(\uparrow)}({k}_{3}) \alpha_{\tau}^{(\downarrow)}({k}_{4}).
\nonumber
\ea
is the bare vertex function. The factor $M(k_1,k_2;k_3,k_4)$ of the scattering element play a crucial role in expressing multiorbital and multi-sublattice interference effects. \cite{kuroki-09prb224511,thomale-11prl187003,kiesel-12prb121105}

\subsection{Perturbative renormalization group}
In the first step of the calculation, we integrate out the high energy degrees of freedom and compute  the various vertex operators that enter the low energy effective action.  The only important vertex is  the Cooper channel, $\Gamma_{\sigma\sigma^{\prime}}(n,\hat{\textbf k};n^{\prime},\hat{\textbf k}^{\prime})$, which is the amplitude for scattering a pair of electrons with spin polarization $\sigma$ and $\sigma^{\prime}$ from crystal momenta $\hat{\textbf k}^{\prime}$ and $-\hat{\textbf k}^{\prime}$ on the Fermi surface corresponding to band $n^{\prime}$, to $\hat{\textbf k}$ and $-\hat{\textbf k}$ on the Fermi surface corresponding to band $n$ while maintaining their spins. This calculation can be carried out perturbatively in powers of $U$ with the result
\begin{eqnarray}
&\Gamma_{\uparrow\downarrow}(\hat{k};\hat{k}^\prime)&=U\Gamma^{(1)}_{\uparrow\downarrow}(\hat{k};\hat{k}^\prime)
+U^2\Gamma^{(2)}_{\uparrow\downarrow}(\hat{k};\hat{k}^\prime)+\ldots
\label{effint1}
\end{eqnarray}
where $\hat{k}\equiv (n,\hat{\bf k})$, $-\hat{k}\equiv (n,-\hat{\bf k})$, $\hat{k}^\prime\equiv (n^\prime,\hat{\bf k}^\prime)$,
\be
\Gamma^{(1)}_{\uparrow\downarrow}(\hat{k};\hat{k}^\prime)=M(\hat{k},-\hat{k};\hat{k}^\prime,-\hat{k}^\prime),
\ee
\ba
\label{effintupdown}
&&\Gamma^{(2)}_{\uparrow\downarrow}(n,\hat{\textbf k};n^{\prime},\hat{\textbf k}^\prime)=\Gamma^{(2)}_{\downarrow\uparrow}(-\hat{\textbf k},n;-\hat{\textbf k}^\prime,n^{\prime})\nonumber\\
&&=-\nu^2\sum_{m,m^\prime}\int \frac{d^{2}\textbf{p}}{(2\pi)^{2}}\Bigg\{
\frac{f[\xi_{\downarrow}(m^{\prime},\textbf{p})]-f[\xi_{\uparrow}(m,\textbf{p}+\hat{\textbf k}
+\hat{\textbf k}^\prime)]} {\xi_{\downarrow}(m^{\prime},\textbf{p})-\xi_{\uparrow}(m,\textbf{p}+\hat{\textbf k}
+\hat{\textbf k}^\prime)}\nonumber\\
&&\quad\times M[(n,\hat{\textbf k}),(m^{\prime},\textbf{p});\,(m,\textbf{p}+\hat{\textbf k}+\hat{\textbf k}^\prime),(n^\prime,-\hat{\textbf k}^\prime)]\\
&&\quad\times M[(m,\textbf{p}+\hat{\textbf k}+\hat{\textbf k}^\prime),(n,-\hat{\textbf k});\,(n^\prime,\hat{\textbf k}^\prime),(m^\prime,\textbf{p})] \Bigg\},
\nonumber
\ea
where $\xi_{\sigma}(n,\textbf{p})\equiv\epsilon_{\sigma}(n,\textbf{p})-\mu$ is the single-particle energy measured from the Fermi level $\mu$, $\nu$  is the volume of the unit cell, $f(\xi)$ is the Fermi function (actually, the step function because we work in zero temperature), and the integral is performed over the first Brillouin zone.

\begin{figure}[t]
\subfigure[]{
\includegraphics[scale=0.8]{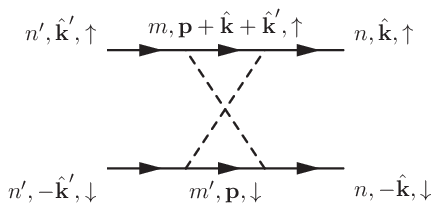}}\\
\subfigure[]{
\includegraphics[scale=0.8]{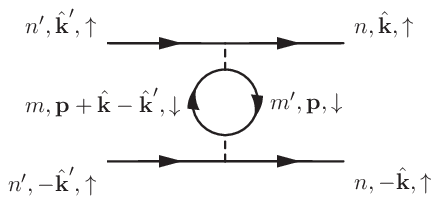}}
\caption{Second-order effective interaction vertices in the Cooper channel.
 (a) Opposite spins: $\Gamma_{\uparrow\downarrow}^{(2)}(n,\hat{\textbf k};n^{\prime},\hat{\textbf k}^{\prime})$.
(b) Like spins: $\Gamma_{\uparrow\uparrow}^{(2)}(n,\hat{\textbf k};n^{\prime},\hat{\textbf k}^{\prime})$. $\Gamma_{\downarrow\uparrow}^{(2)}$ and $\Gamma_{\downarrow\downarrow}^{(2)}$ are obtained by inverting all spins in (a) and (b), respectively.}
\label{diagram}
\end{figure}

For like spins, there is no first order term, $\Gamma^{(1)}_{\sigma\sigma}=0$, since the on-site Hubbard interaction operates only between antiparallel spins, while
\ba
\label{effintupup}
&&\Gamma^{(2)}_{\uparrow\uparrow}(n,\hat{\textbf k};n^{\prime},\hat{\textbf k}^\prime)\nonumber\\
&&=\nu^2\sum_{m,m^\prime}\int \frac{d^{2}\textbf{p}}{(2\pi)^{2}}\Bigg\{
\frac{f[\xi_{\downarrow}(m^{\prime},\textbf{p})]-f[\xi_{\downarrow}(m,\textbf{p}+\hat{\textbf k}
-\hat{\textbf k}^\prime)]} {\xi_{\downarrow}(m^{\prime},\textbf{p})-\xi_{\downarrow}(m,\textbf{p}+\hat{\textbf k}
-\hat{\textbf k}^\prime)}\nonumber\\
&&\quad\times M[(n,\hat{\textbf k}),(m^{\prime},\textbf{p});\,(n^\prime,\hat{\textbf k}^\prime),(m,\textbf{p}+\hat{\textbf k}-\hat{\textbf k}^\prime)]\\
&&\quad\times M[(n,-\hat{\textbf k}),(m,\textbf{p}+\hat{\textbf k}-\hat{\textbf k}^\prime);\,(n^\prime,-\hat{\textbf k}^\prime),(m^\prime,\textbf{p})] \Bigg\},
\nonumber
\ea
and $\Gamma_{\downarrow\downarrow}^{(2)}(\hat{k};\hat{k}^\prime)$ is defined correspondingly.  Figure \ref{diagram} shows the Feynman diagrams corresponding to Eqs. (\ref{effintupdown}) and (\ref{effintupup}). The momentum transfer is $\hat{\textbf k}+\hat{\textbf k}^\prime$ for opposite and $\hat{\textbf k}-\hat{\textbf k}^\prime$ for like spins. When there is only a single band (per spin), further simplification follows:
\be
\Gamma_{\sigma,\pm\sigma}^{(2)}(\hat{\textbf k};\hat{\textbf k}^{\prime}) = \pm \,\chi_{-\sigma,\mp\sigma}(\hat{\textbf k}\mp\hat{\textbf k}^\prime),
\ee
where
\be
\chi_{\sigma\sigma^{\prime}}(\textbf{q}) \equiv \nu^2\int \frac{d^{2}\textbf{p}}{(2\pi)^{2}}\Bigg\{
\frac{f[\xi_{\sigma}(\textbf{p})]-f[\xi_{\sigma^{\prime}}(\textbf{p}+\textbf{q})]} {\xi_{\sigma}(\textbf{p})-\xi_{\sigma^{\prime}}(\textbf{p}+\textbf{q})}\Bigg\},
\ee
is the usual particle-hole susceptibility.

The second stage of the RG calculation is most efficiently carried out in a basis that diagonalizes $\Gamma$, in which  each component renormalizes independently of the others.  The eigenvalue equation for pairing with antiparallel spins assumes the form of an integral equation:
\begin{equation}
\begin{split}
\label{eigen}
\sum_{n^{\prime}}&\int\!\frac{d\hat{\textbf k}_{n^{\prime}}^{\prime}}{(2\pi)^{2}v(n^{\prime},\hat{\textbf k}_{n^{\prime}}^{\prime})}\Gamma_{\uparrow\downarrow}(n,\hat{\textbf k}_{n};n^{\prime},\hat{\textbf k}_{n^{\prime}}^{\prime})
\psi_{\alpha}(n^{\prime},\hat{\textbf k}_{n^{\prime}}^{\prime})\\
&=\lambda_{\alpha} \psi_{\alpha}(n,\hat{\textbf k}_{n})\,,
\end{split}
\end{equation}
where $v(n,\hat{\textbf k}_{n})\equiv \partial\epsilon(n,\textbf{k})/\partial \textbf{k}$ denotes the Fermi velocity at $(n,\hat{\textbf k}_{n})$. The analogous expression for like-spin pairing is obtained in terms of $\Gamma_{\sigma\sigma}$. The integration is over the portion of the Fermi surface belonging to each band, and we have introduced new subscripts $n$ and $n^{\prime}$ in $\hat{\textbf k}_{n}$ and $\hat{\textbf k}_{n^{\prime}}^{\prime}$ to emphasize the bands they originate from. In cases in which SU(2) spin rotational symmetry is preserved, all even eigenfunctions of $\Gamma_{\uparrow\downarrow}$ correspond to spin-singlet and all odd eigenfunctions of both $\Gamma_{\uparrow\downarrow}$ and $\Gamma_{\uparrow\uparrow}=\Gamma_{\downarrow\downarrow}$ correspond to spin-triplet pairing. Notice that $1/v$ accompanies the integration measure. The eigenfunctions in Eq. (\ref{eigen}) satisfy the normalization condition
\begin{equation}
\sum_{n}\int\!\frac{d\hat{\textbf k}_{n}}{(2\pi)^{2}v(n,\hat{\textbf k}_{n})}  \psi_{\alpha}^{\ast}(n,\hat{\textbf k}_{n})\psi_{\beta}(n,\hat{\textbf k}_{n}) = \delta_{\alpha\beta}
\end{equation}
and the completeness relation
\ba
\sum_{\alpha} \psi_{\alpha}(n,\hat{\textbf k}_{n}) \psi_{\alpha}^{\ast}(n^{\prime},\hat{\textbf k}_{n^{\prime}}^{\prime}) = (2\pi)^{2}v(n,\hat{\textbf k}_{n}) \delta(\hat{\textbf k}_{n}-\hat{\textbf k}_{n^{\prime}}^{\prime})\,\delta_{n n^{\prime}} \,.
\nonumber
\ea

For small $U$, the first-order contribution to the Cooper channel vertex for opposite-spin pairing, $U\Gamma_{\uparrow\downarrow}^{(1)}$, is parametrically large compared to the second-order contribution.  As $\Gamma^{(1)}_{\uparrow\downarrow}$ is positive semidefinite, the space of all possible pairing solutions can be divided into two subspaces---that in which $\Gamma^{(1)}_{\uparrow\downarrow}$ has positive eigenvalues (which would be the space of the ``conventional $s$-wave'' states if $U$ were attractive) and that which is annihilated by $\Gamma^{(1)}_{\uparrow\downarrow}$ (which we will refer to as ``unconventional'').  In analyzing higher order contributions to $\Gamma_{\uparrow\downarrow}$, we must always project onto the unconventional subspace---this will be implicit in all further discussion.

All negative eigenvalues grow under the second stage of RG.  We will adopt the convention that $\lambda_n \leq \lambda_{n+1}$, so $\lambda_0$ is the most negative, and hence most relevant eigenvalue. The RG procedure can be iterated until the most relevant coupling grows to be of order 1, which occurs at an energy scale $\sim W \exp(-1/|\lambda_{0}|)$, where $W$ is the bandwidth. \cite{raghu-10prb224505} This energy scale is identified with the superconducting transition temperature, and the symmetry of the eigenvector(s) corresponding to $\lambda_{0}$ is the pairing symmetry. Solving the BCS gap equation using this interaction gives a pairing gap that has exactly the same momentum dependence as one of the pair wave-functions $\psi_{0}^{(n)}(\hat{\textbf k}_{n})$. Moreover, $\psi_{0}^{(n)}(\hat{\textbf k}_{n})$ must transform according to an irreducible representation of the point group, and the symmetry of the pairing gap (i.e., the pairing symmetry) can be classified in the same manner. Note that different harmonics can contribute to the given form factor in its corresponding symmetry sector.

Finally, we comment on practical methods for diagonalizing $\Gamma$. This is performed numerically, after discretizing Eq. \ref{eigen} so that it becomes a matrix equation. We can take two different approaches in the discretization. Most straightforwardly, we can do this by discretizing the Fermi  surface in terms of a large set of patches, for which various schemes have already been developed within the functional renormalization group (fRG). \cite{halboth-00prb7364,honerkamp-01prb035109,wang-09prl047005,thomale-11prl187003,metzner-12rmp299,hur20091452} Alternatively, we can discretize the whole first Brillouin zone, but only keep states whose single-particle energy lies within a small energy window $\Omega$ from the Fermi surface. This requires transforming the integration measure in Eq. (\ref{eigen}) into a summation over the momenta within the energy window:
\begin{equation}
\begin{split}
\label{discretization2}
\int\!&\frac{d\hat{\textbf k}_{n}}{(2\pi)^{2}v(n,\hat{\textbf k}_{n})} = \frac{1}{2\Omega}\int_{-\Omega}^{\Omega}\!d\xi
\int\!\frac{d\hat{\textbf k}_{n}}{(2\pi)^{2}v(n,\hat{\textbf k}_{n})}\\
&\approx\frac{1}{2\Omega} \int_{{}_{|\xi_{n}\!(\textbf{k})|<\Omega}}\!\!\frac{d^{2}\textbf{k}}{(2\pi)^{2}}
 \approx\frac{1}{2\Omega N
 \nu\!\!}\!\!\sum_{\substack{\textbf{k}\\|\xi_{n}\!(\textbf{k})|<\Omega}}\,,
\end{split}
\end{equation}
where $N$ is the number of unit cells in the system.

The first method is more accurate for a fixed number of discrete $k$-points (fixed size matrix). At the same time, implementing it poses a challenge, as a detailed analysis of the equation describing the Fermi surface is required to divide it into different segments and determine the weight for each. The second method is easier to implement, but requires a much larger number of points to ensure the same level of accuracy. Eventually, we have employed the first method, taking advantage of the expertise existent from the fRG schemes.

\subsection{Pairing from repulsive interactions}
\label{general}
Before analyzing specific problems, we discuss the considerations which lead to negative eigenvalues of the Cooper channel vertex $\Gamma$.

In many circumstances, the second-order Cooper channel vertex is a quantity with a fixed sign regardless of its band indices or momentum arguments. When $t_{\tau\tau^{\prime}}^{( \sigma)}(\textbf{k})=t_{\tau\tau^{\prime}}^{( -\sigma)}(-\textbf{k})^{\ast}$ (time-reversal symmetry) in Eq. (\ref{model}), $\alpha_{\tau}^{\sigma}(k)=\alpha_{\tau}^{-\sigma}(-k)^{\ast}$ is satisfied. Then, the two bare vertex functions in Eq. (\ref{effintupdown}) are complex conjugates of each other and it follows that $\Gamma_{\sigma,-\sigma}^{(2)}>0$. Similarly, $t_{\tau\tau^{\prime}}^{( \sigma)}(\textbf{k})=t_{\tau\tau^{\prime}}^{( \sigma)}(-\textbf{k})^{\ast}$ (all hopping amplitudes are real) implies $\Gamma_{\sigma\sigma}^{(2)}<0$. \footnote{These conditions are somewhat more stringent than the requirement on the single-particle spectrum for weak-coupling superconducting instability to be possible, i.e., $\epsilon_{\sigma}(\textbf{k})=\epsilon_{-\sigma}(-\textbf{k})$ or $\epsilon_{\sigma}(\textbf{k})=\epsilon_{\sigma}(-\textbf{k})$, because any $\textbf{k}$-dependent unitary transformation in the sublattice space would give the same dispersion.} In such cases, which includes all examples we have studied except for opposite-spin pairing in the presence of an SDW background, negative eigenvalues of the positive quantity $\Gamma_{\uparrow\downarrow}^{(2)}$ arise from large off-diagonal elements, leading to sign-changing pair wave-functions. On the other hand, to obtain a negative eigenvalue out of the negative quantity $\Gamma_{\sigma\sigma}^{(2)}$, it is desirable to have a pair wave-function with the same sign between two momenta whenever the matrix element connecting them is large, together with the sign change mandated by fermion antisymmetry.

The features that characterize the states with the most negative eigenvalues are then clear. (1)  They involve sign-changing order parameters which are orthogonal (averaged along the Fermi surface) to all ``conventional'' $s$-wave states.  (2)  The structure of the favored superconducting gap along the Fermi surface can be inferred in large part from a catalog of wave vectors, ${\textbf Q}$, at which $\Gamma^{(2)}$ is large;  in general, for anti-parallel spins, the gap will have the opposite sign at any two points on the Fermi surface for which $\hat{\textbf k}+\hat{\textbf k}^\prime\approx{\textbf Q}$, and for parallel spins, to the extent it is consistent with the antisymmetry required by Fermi statistics, the gap will have a uniform sign for any two points for which  $\hat{\textbf k}-\hat{\textbf k}^\prime\approx{\textbf Q}$.  (3)  The most important portions of the Fermi surface are either those in which this approximate ``nesting'' condition is satisfied over a substantial region of the Fermi surface, or in which the Fermi velocity is small (density of states is large).  Portions of the Fermi surface with a relatively small integrated density of states generally play  little role in either determining the structure of the gap function, or  the magnitude of the eigenvalue, $\lambda_0$.

Notice that as long as spin rotation symmetry is unbroken, both spin-singlet and spin-triplet pairing can be derived from $\Gamma_{\uparrow\downarrow}$---even parity solutions are singlet and odd parity solutions are triplet. For magnetically ordered states that break spin rotation symmetry, the eigenfunctions of $\Gamma_{\sigma\sigma}$ must be analyzed separately.

\section{Model Problems}
\label{bandstructures}
\subsection{Square-lattice Hubbard model}

We have applied the above theoretical framework to a variety of systems based on the 2D square lattice. The common starting point is the repulsive Hubbard model with nearest- and next-nearest-neighbor hopping amplitudes ($t$ and $t^{\prime}$) and uniform on-site energy (which is set to zero). The energy dispersion in the non-interacting limit is thus
\begin{equation}
\epsilon(\textbf{k}) = - 2t(\cos k_{x} + \cos k_{y}) - 4t^{\prime }\cos k_{x}\cos k_{y} \, .
\end{equation}
In Ref.~\onlinecite{raghu-10prb224505}, various cuts through the phase diagram of this model were computed for weak repulsive $U$. For both $t^{\prime}=0$ and $t^{\prime}=-0.3t$, the ground state exhibits superconductivity with the $d_{x^{2}-y^{2}}$ symmetry for a broad range of electron density near half filling.

\begin{figure}[t]
\includegraphics[scale=0.28]{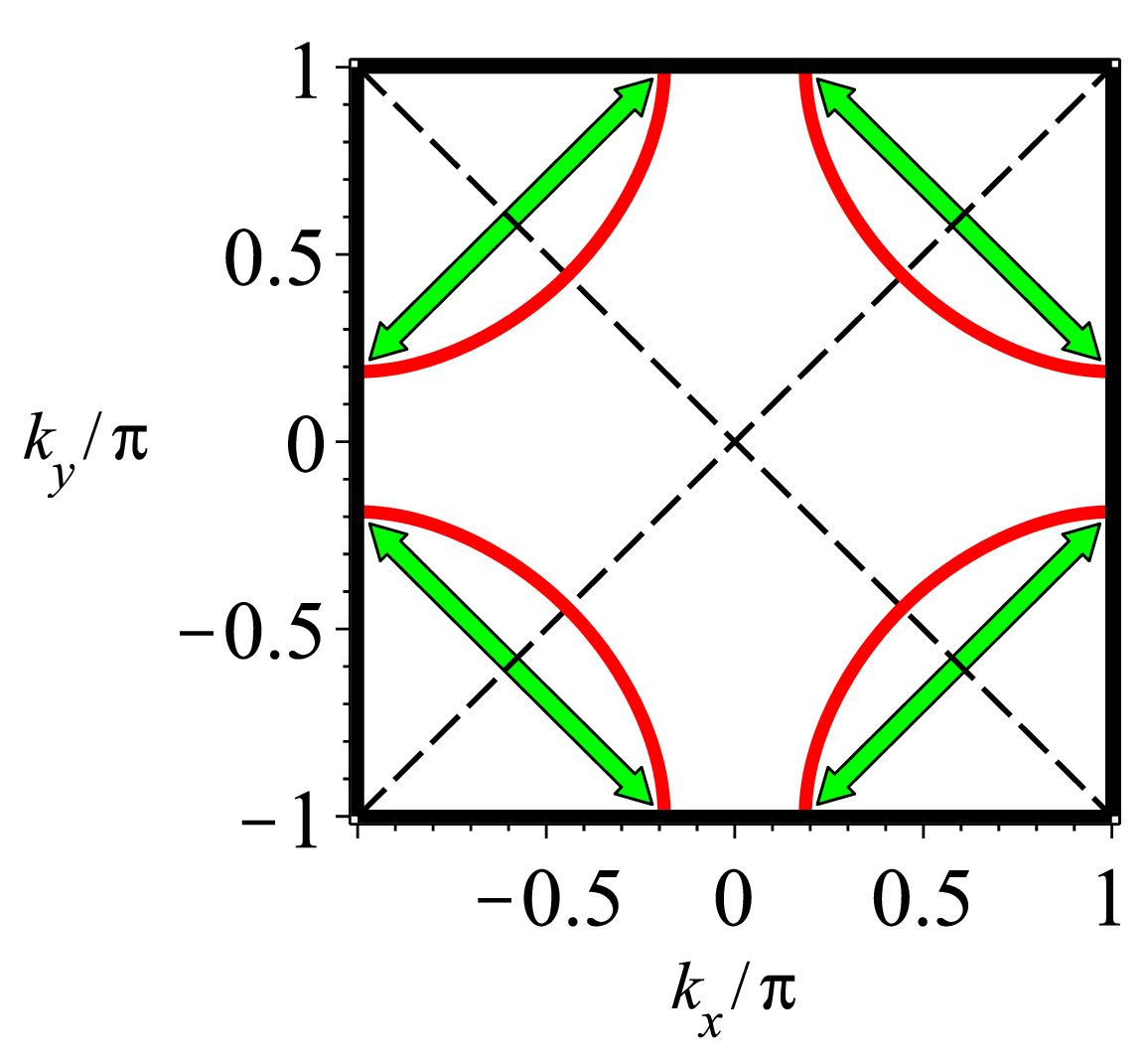}
\caption{The Fermi surface of the ``parent Hubbard model'' for $t^{\prime}=-0.3t$ at half filling ($n_{\textmd{el}}=1$). The green arrows represent the dominant scattering processes, and the dashed lines mark the nodes of the $d_{x^{2}-y^{2}}$-wave gap.}
\label{fs0}
\end{figure}

Before discussing more complicated models, let us review the origins of the robust $d_{x^{2}-y^{2}}$-wave superconductivity in this model. No multiorbital or multi-sublattice effects are present [$M(k_1,k_2,k_3,k_4)=1$].  Thus, to first order in $U$, $\Gamma_{\uparrow\downarrow}$ is momentum independent and repulsive.  As a consequence,  all candidate pair wave functions, $\psi_\alpha(\hat{\textbf k})$, are constrained to be orthogonal to the trivial $s$-wave solution, $\psi_s(\hat{\textbf k})\propto 1$. The second-order contribution is $-U^{2}\chi(\textbf{q}=\hat{\textbf k} + \hat{\textbf k}^{\prime})>0$ (spin indices are dropped because there is no spin dependence), which for electron density not too different from $n_{\textmd{el}}=1$ per site, is somewhat peaked for ${\textbf Q}=(\pi,\pi)$. [$\chi({\textbf q})$ has a logarithmic divergence at ${\textbf q}={\textbf Q}$ under fine-tuned circumstances when the Fermi surface passes through the Van Hove points at $(\pi,0)$ and $(0,\pi)$.]  Moreover, the density of states is maximal near $(0,\pm\pi)$ and $(\pm\pi,0)$, while minimal where diagonals of the square Brillouin zone intersect the Fermi surface. Thus, the  second-order induced interaction vertex is the largest between the region near $(0,\pm\pi)$ and that near $(\pm\pi,0)$, precisely where the density of states is largest. For $d_{x^2-y^2}$ pairing, the gap changes sign in just the right way to take full advantage of this strong effect, while the  associated nodes intersect the Fermi surface exactly where the density of state is the smallest, and hence the associated loss in condensation energy is smallest (see Fig. \ref{fs0}).

\subsection{Ordered translationally invariant states}

With translation symmetry unbroken, we are still dealing with a single band problem, for which $M=1$ and hence $\Gamma^{(2)}$ simply equals the usual particle-hole susceptibility $\chi$ (up to a sign).  Thus, the only differences with the problem already analyzed come from the effects of the ${\textbf k}$-space structure of $\chi$ and the shape of the Fermi surface produced by the various changes in the bandstructure.

\subsubsection{Nematic phase}

The nematic phase breaks the $D_{4}$ point group symmetry of the square lattice down to $D_{2}$.
Time-reversal, inversion, spin SU(2), and lattice translational symmetries are unbroken.

\begin{figure}[t]
\subfigure[]{
\includegraphics[scale=0.8]{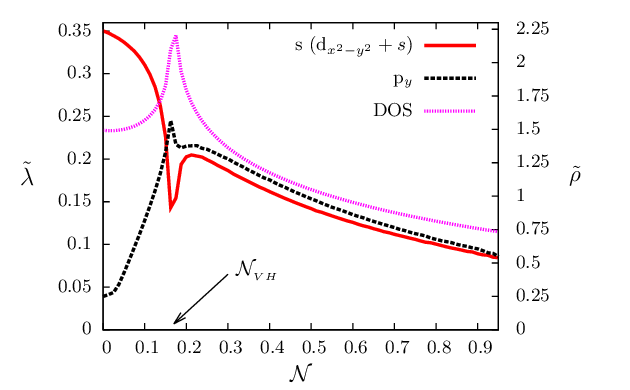}}\\
\subfigure[]{
\includegraphics[scale=0.8]{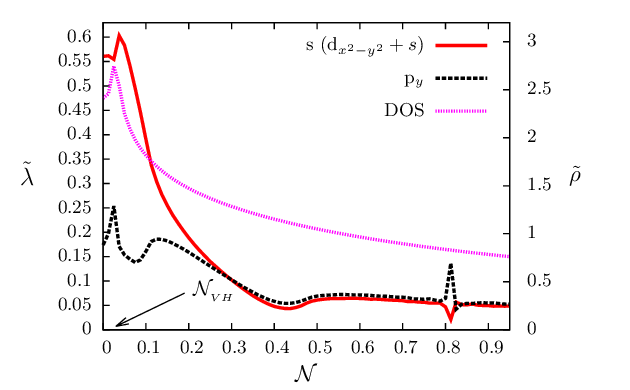}}
\caption{Pairing strengths and the density of states ($y$ axis to the right) as functions of nematicity with $t^{\prime}=-0.3t$. (a) $n_{\textmd{el}}=1.0$ (half filling).
(b) $n_{\textmd{el}}=0.8$. ${\cal N}_{VH}$ denotes the critical value of nematicity at which the Fermi surface changes from being closed and hole-like to open and quasi-1D. In (b), the singularity at ${\cal N}\approx0.81$ marks the special point at which the Fermi surface is perfectly flat.}
\label{nematicpairing}
\end{figure}

Nematic order can be realized by assigning different hopping amplitudes to $x$ and $y$ directions, i.e., $t_{x} = t(1-{\cal N})$ and $t_{y} = t(1+{\cal N})$, where the nematic order parameter ${\cal N}$ ($0\le{\cal N}\le1$) controls the anisotropy of the system. The band structure is now modified to
\begin{equation}
\begin{split}
\epsilon(\textbf{k}) =& -2t\big[(1-{\cal N})\cos k_{x} + (1+{\cal N}) \cos k_{y} \big]\\
&- 4t^{\prime }\cos k_{x}\cos k_{y} \,.
\end{split}
\label{nematic}
\end{equation}
Because the nematic phase breaks the point group symmetry, certain pairing channels for the undistorted system are no longer distinguishable by symmetry.  For example, $d_{x^{2}-y^{2}}$ and $s$ are mixed in the nematic state, as are $g_{xy(x^{2}-y^{2})}$ and $d_{xy}$. The $p$-wave channel, which forms a two-dimensional representation when the nematic order parameter vanishes, splits into distinct $p_{x}$ and $p_{y}$ states.

\begin{figure}[t]
\subfigure[]{
\includegraphics[scale=0.21]{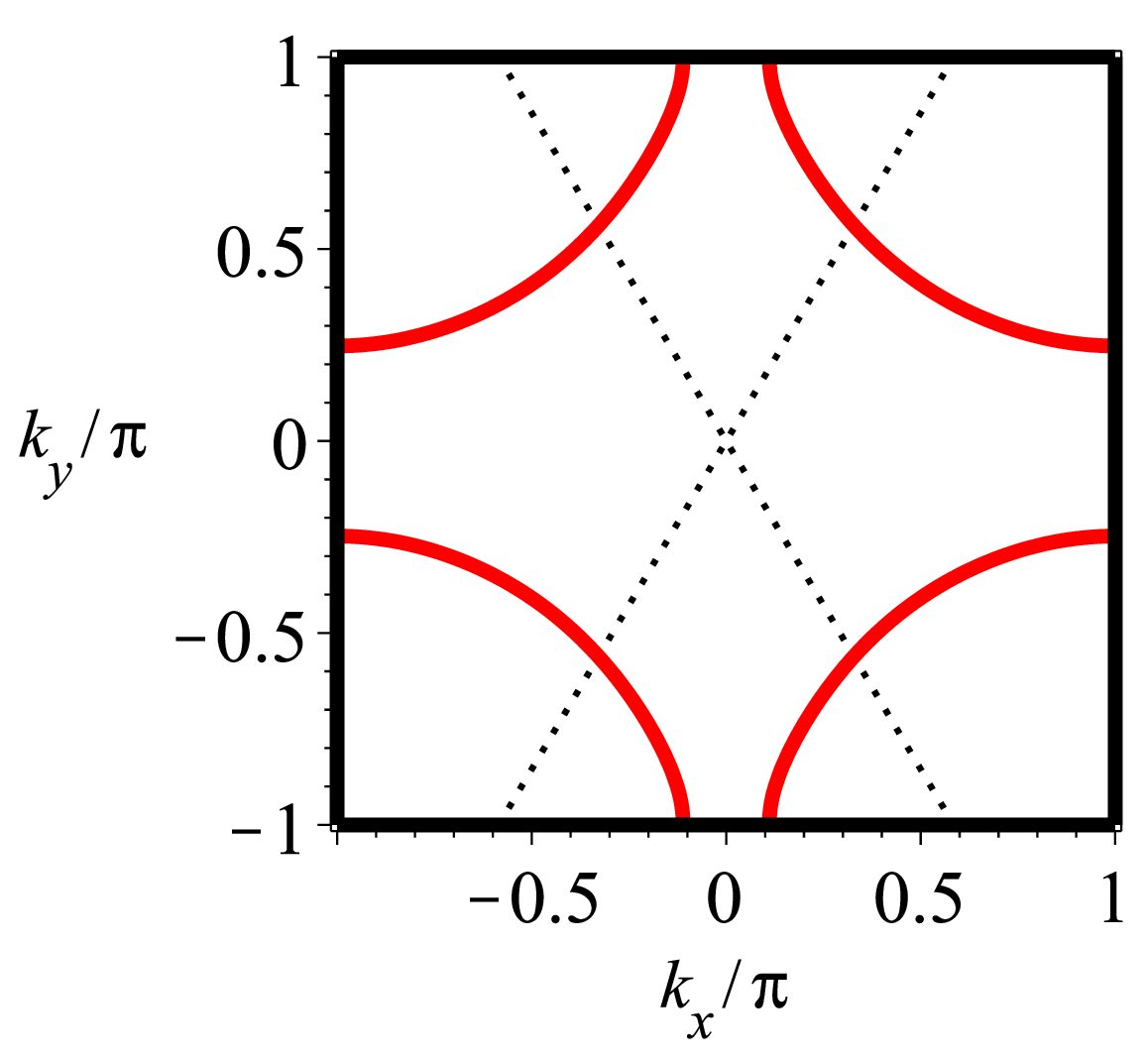}}
\subfigure[]{
\includegraphics[scale=0.21]{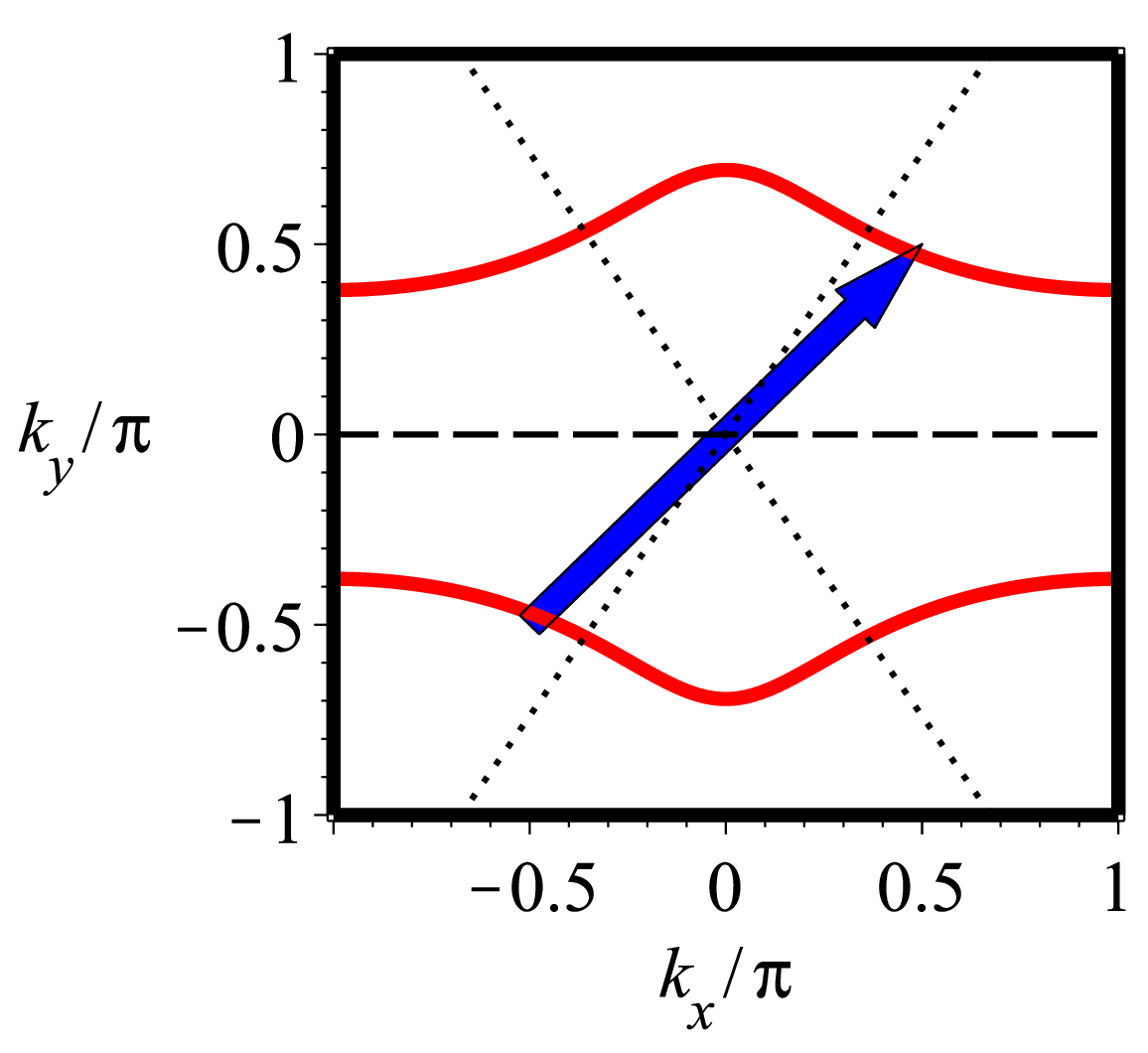}}
\caption{Fermi surfaces in the presence of the nematic background for $t^{\prime}=-0.3t$ and $n_{\textmd{el}}=1$ (${\cal N}_{VH}\approx0.17$).
(a) Small anisotropy (${\cal N}=0.1$): the dotted lines denote the accidental nodes of the $(d+s)$-wave gap. (b) Large anisotropy (${\cal N}=0.4$): the dashed line is the symmetry-required nodal line of the $p_{y}$-wave gap, while the dotted lines denote approximate locations of the accidental nodes for both the $(d+s)$- and $p_{y}$-wave gap functions. The blue arrow represents the approximate nesting vector $(\pi,2k_{F})$, where $2k_{F}=\pi n_{\textmd{el}}$.}

\label{nematicFS}
\end{figure}

Figure \ref{nematicpairing} shows pairing strengths and density of states for $t^{\prime}=-0.3t$ at half-filling ($n_{\textmd{el}}=1$) and at $n_{\textmd{el}}=0.8$ as a function of ${\cal N}$. Here and henceforth, the quantities plotted will be
\be
\tilde{\lambda}\equiv \frac{W^{2}|\lambda|}{U^{2}}, \qquad \tilde{\rho} \equiv \rho W,
\ee
where the bandwidth is given by $W=8t$.

Figure \ref{nematicFS} shows the Fermi surfaces for small and large anisotropy. We see from Fig.~\ref{nematicpairing} that the dominant pairing instability changes from $s$ (inherited from the $d_{x^{2}-y^{2}}$ wave for the original square lattice) to $p_{y}$ as ${\cal N}$ is increased. There are two notable points in the behavior of pairing strengths as ${\cal N}$ is varied. First, the sharp feature in the vicinity of the Van Hove singularity, and second, near-degeneracies between different pairing channels which persists for a broad range of ${\cal N}$ where the system is quasi-one-dimensional.
In the range of ${\cal N}$ and $n_{\textmd{el}}$ considered here, the Fermi surface passes through the Van Hove point at $\textbf{k}=(0,\pi)$ at a critical value ${\cal N}={\cal N}_{VH}$, (${\cal N}_{VH}\approx0.17$ for $n_{\textmd{el}}=1$, and ${\cal N}_{VH}\approx0.03$ for $n_{\textmd{el}}=0.8$). For ${\cal N}<{\cal N}_{VH}$, the Fermi surface is closed around $(\pi,\pi)$, whereas for ${\cal N}>{\cal N}_{VH}$, it is open along the $x$ direction. The pairing strengths, $\lambda_a$, in the various channels are continuous, but nonanalytic function of ${\cal N}$ at ${\cal N}_{VH}$.

By comparing $\lambda({\cal N})$ with $\lambda(0)$ in Fig.~\ref{nematicpairing}, we can address the question of whether nematicity competes with or enhances superconductivity.  We see the common feature for both $n_{\textmd{el}}=1$ and $n_{\textmd{el}}=0.8$ that singlet ($d+s$)-wave superconductivity is at first suppressed by increasing nematicity, but the behavior is abruptly inverted at the critical value ${\cal N}_{VH}$, and beyond this point, the pairing strength increases as a function of ${\cal N}$ and then drops again past a local maximum. For $n_{\textmd{el}}=0.8$, where ${\cal N}_{VH}$ is small, this local maximum corresponds to an enhancement of superconductivity due to nematicity, i.e., Max$[\lambda({\cal N})]>\lambda(0)$. (Recall, $T_c$ depends exponentially on $\lambda$, so what may appear to be a small relative effect in $\lambda$ corresponds, in the weak-coupling limit, to a large effect in $T_c$.) On the other hand, for $n_{\textmd{el}}=1$, where ${\cal N}_{VH}$ is relatively larger, the pairing strength is uniformly weaker for non-zero ${\cal N}$ than for ${\cal N}=0$.

In the limit of large ${\cal N}$, the nematic band structure corresponds to that of a quasi-1D conductor.
There are a few salient features of superconductivity in this limit that warrant mention.  Firstly, for most of the range of ${\cal N}>{\cal N}_{VH} $, there is a remarkable near degeneracy of the singlet $(d+s)$-wave and the triplet $p$-wave pairing channels.  Moreover, both the corresponding pair wave functions   are more structured than expected---on each open segment of the Fermi surface, they both have a pair of ``accidental nodes'' that  are not required by symmetry.  Indeed, the near degeneracy of the two states arises from the fact that the gap structure looks almost the same on each open Fermi surface, with the only significant difference being the sign change in going from one Fermi surface to the other in the $p$-wave case.  (While the actual gap structure is generally quite complicated, a caricature of the state is that $\Delta({\textbf k})=\Delta_0\big\{\cos[(1+\phi)k_{x}] - \cos[(1-\phi)k_{y}]\big\}$ for the singlet state and $\Delta({\textbf k})=\Delta_0\sin(k_y)[ \cos(k_x)-\delta]$ for the triplet state, where both $\phi$ and $\delta$ depend on the precise shape and position of the Fermi surface, but are both relatively small.)

At an intuitive level, we can think of the accidental nodes in the $(d+s)$-wave state as reflecting the smooth evolution of this state from a pure $d$-wave parent state in the isotropic lattice (${\cal N}=0$);  while there is no symmetry mandating it, this state remains largely $d$-wave-like in character.  This  underscores the existence of a single ``mechanism'' underlying the $d$-wave superconductivity of the square lattice and unconventional $(d+s)$-wave pairing in quasi 1D.  However, this unification is still broader---the nearly identical nodal structure and pairing strength in the triplet channel carries with it the implication that the mechanism of pairing can be essentially identical, independent of the symmetry of the order parameter!  (This connection has also been observed \cite{srisconferencepaper} in the context of weak-coupling calculations of models with the band structure of Sr$_2$RuO$_4$.)

The origin of both the accidental near degeneracy of two symmetry-distinct superconducting states and of the accidental nodes can be understood simply in the quasi-1D limit.
As confirmed by our study, $\chi$ in a quasi-1D conductor is peaked at $(\pi,2k_{F})$, where $2k_{F}=\!\pi n_{\textmd{el}}$ is the average separation between the two branches of the open Fermi surface [see Fig. \ref{nematicFS}(b)]. Thus, in looking for the most negative eigenvalues of the positive quantity $\Gamma_{\uparrow\downarrow}^{(2)}(\hat{\textbf k};\hat{\textbf k}^\prime)=-\chi(\hat{\textbf k}+\hat{\textbf
k}^{\prime})$, the most significant processes are those with $\hat{\textbf k}+\hat{\textbf k}^\prime$ close to $(\pi,\pm 2k_{F})$, i.e., the nesting vector shown in Fig. \ref{nematicFS}(b).  Note that this refers to particle-hole excitations from one branch of the Fermi surface to the other; the corresponding process in which a particle pair is scattered from $({\bf \hat k}^\prime,-{\bf \hat k}^\prime)$ to $({\bf \hat k},-{\bf \hat k})$ requires  ${\bf \hat k}$ and ${\bf \hat k}^\prime$ be on the same branch of the Fermi surface. That interbranch scattering has a subdominant effect implies that the relative sign of the pairing gap on the two is relatively unimportant;  this is responsible for the near-degeneracy of the $d+s$ and $p_y$ states, and between $d_{xy}$ and $p_x$ states. (For the case of the dominant orders $d+s$ and $p_y$, see Fig.~\ref{nematicpairing}.) Moreover, $\hat{\textbf k}+\hat{\textbf k}^{\prime}\approx(\pi,\pm2k_{F})$ requires $\hat{\textbf k}$ and $\hat{\textbf k}^{\prime}$ to be not only on the same branch but also in the same quadrant of the Brillouin zone. This results in an accidental node in each quadrant, as depicted in Fig. \ref{nematicFS}(b).

Notice that the superconducting properties of the repulsive Hubbard model in the presence of strong nematic background is very similar to that of the Bechgaard salts in many aspects: Fermi surface topology, near-degeneracies, and accidental nodes. \cite{bourbonnais08,doiron-leyraud-09prb214531} There, given that the superconducting phase is induced by electronic interaction, the close competition between $d$-wave- and $p$-wave-type phases is to be expected due to the quasi-one-dimensional nature of the band structure. As the competing superconducting states are located in different spin channels, a way to test the applicability of these ideas would be to apply a weak Zeeman field to the material, which would favor the spin-aligned $p$-wave order over the $d$-wave one.

\subsubsection{Nematic spin nematic phase}

\begin{figure}[t]
\subfigure[]{
\includegraphics[scale=0.8]{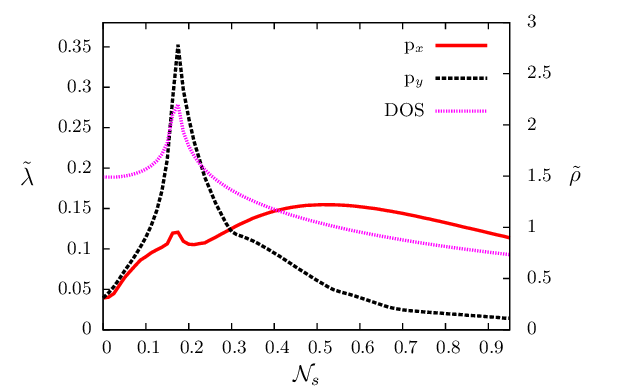}}\\
\subfigure[]{
\includegraphics[scale=0.8]{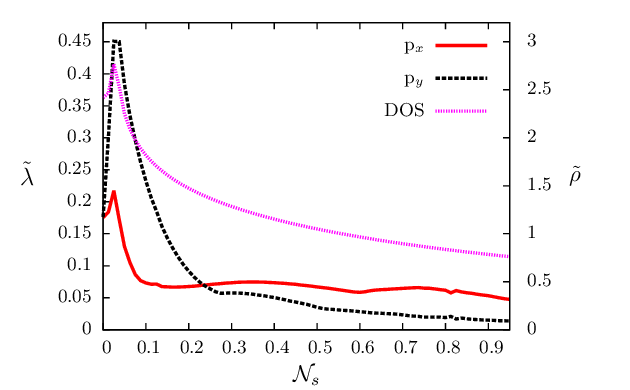}}
\caption{Pairing strengths and the density of states ($y$ axis to the right) as functions of the nematic spin nematic order parameter for $t^{\prime}=-0.3t$. (a) $n_{\textmd{el}}=1.0$ (half filling).
(b) $n_{\textmd{el}}=0.8$. In (b), the (almost invisible) singularity around ${\cal N}_{s}\approx0.81$ is where the Fermi surface is exactly flat.}
\label{snematicpairing}
\end{figure}

Here, we consider a spin-triplet version of nematic order, in which the sign of the nematic distortion  is opposite for spin up and spin down electrons.  Alternatively, this can be thought of as a $d$-wave relative of ordinary ferromagnetism (which could arise in Fermi liquid theory from a sufficiently negative $F_{2}^{a}$).  It has been awkwardly named \cite{oganesyan-03rmp1201} ``nematic spin nematic.''  From the perspective of unconventional superconductivity, the most remarkable thing about this state is that, under appropriate circumstances, it can give rise to a gap structure with a large number of accidental (i.e., unrelated to symmetry) gap nodes.

The nematic spin nematic phase is a nematic phase such that ${\cal N}_{\uparrow}=-{\cal N}_{\downarrow}\equiv{{\cal N}}_{s}$. One immediate consequence is that in the resulting band structure, $\epsilon_{\uparrow}(\textbf{k}) = \epsilon_{\downarrow}(-\textbf{k})$ no longer holds in general.
In the weak-coupling limit, pairing between opposite spins cannot occur, so the only possible superconducting states are $p_{x}$ and $p_{y}$ waves formed by equal-spin pairing. As time reversal followed by a $C_{4}$ rotation remains a good symmetry of the system, spin-up and -down Fermi seas are rotated by 90$^{\circ}$ from each other, and therefore, the $p_{x(y)}$ wave of up spins and the $p_{y(x)}$ wave of down spins are degenerate. Hence, without loss of generality, we hereafter only consider spin-up electrons with ${\cal N}_{s}>0$, and the pairing strengths in this case are shown in Fig.~\ref{snematicpairing}. We see that the pairing symmetry is $p_{x}$ for sufficiently large values of ${\cal N}_{s}$, and otherwise $p_{y}$.

At ${\cal N}_{s}={\cal N}_{1D}\equiv 1+2\frac{t^{\prime}}{t}\cos(\frac{\pi n_{\textmd{el}}}{2})$ (${\cal N}_{1D}=1$ for $n_{\textmd{el}}=1$ and ${\cal N}_{1D}\approx 0.81$ for $n_{\textmd{el}}=0.8$), the Fermi surface for spin up electrons becomes two straight lines given by $k_{y}=\pm n_{\textmd{el}}\pi/2$. In this case, there is a simple mechanism responsible for $p_{x}$-wave superconductivity. In the pairing between up spins, the virtual particle-hole pairs originate from the spin-down Fermi sea [see Eq. (\ref{effintupup}) and Fig. \ref{diagram}(b)], which is perfectly nested for any momentum transfer of the form, $(\pi n_{\textmd{el}},q_{y})$. The second-order interaction $\Gamma_{\uparrow\uparrow}$ is negative in this case. To obtain the most negative eigenvalue, the pair wave function should have the same sign between two points on the spin-up Fermi surface whenever their difference satisfies the nesting condition.

For an arbitrary point on the spin-up Fermi surface, the points shifted by $(\pi n_{\textmd{el}},0)$ and $(\pi n_{\textmd{el}},\pm\pi n_{\textmd{el}})$, respectively, are also on the Fermi surface. As argued above, the pair wave function should have the same sign at all of these points. For the latter two points, which are related to each other by the reflection about the $x$ axis, this is only possible for a $p_{x}$ wave but not a $p_{y}$ wave, and hence $p_{x}$ should be the preferred pairing symmetry. For the former two, it means that the pair wave function has the same sign when translated by $\pi n_{\textmd{el}}$ on each branch of the Fermi surface. This periodicity (in sign), together with the fact that a $p_{x}$-wave is odd under  reflection about the $y$ axis, requires (possibly a large number of) additional nodes. One can show that when $n_{\textmd{el}}/2 = a/b$, where $a$ and $b$ are two integers that are relatively prime, $2b$ total number of evenly spaced nodes should exist in each branch. Notice that having nodes is unfavorable for pairing in general because of reduced condensation energy. Hence, we expect that larger $b$ would result in lower $T_{c}$ at ${\cal N}_{s}={\cal N}_{1D}$. This is indeed seen in Fig.~\ref{snematicpairing}, as the pairing strength in the dominant $p_{x}$ channel is significantly smaller for $n_{\textmd{el}}=0.8$ ($b=5$) than for $n_{\textmd{el}}=1$ ($b=2$).

\subsubsection{Orbital current-loop order}

A fascinating orbital current-loop ordered state has been proposed \cite{varma06prb155113} to account for many of the features of the pseudo-gap of the cuprates.  Among all proposals for broken symmetry states in the pseudo-gap, this is unique in that it breaks both time-reversal and inversion symmetries.  A consequence of this is that the perfect nesting in the particle-particle channel responsible for the Cooper instability is absent in this state, \cite{barkeshli-13prb140402} which in turn implies that for sufficiently weak coupling, this order is incompatible with superconductivity. In numerical approaches designed to treat stronger coupling regime for finite size systems, it has likewise been an ongoing challenge to detect orbital loop current order in Hubbard models for cuprate superconductors. \cite{greiter-07prl027005,nishimoto-09prb205115,weber-09prl017005} From a weak coupling viewpoint, no other order considered to date so unambiguously ``competes'' with superconductivity.

\subsection{$(\pi,\pi)$-density-wave orders}
\label{pipiDW}

We now move on to states that break the translational symmetry of the underlying square lattice. Here, there is more than one band, so the bare interaction vertex $M$ is non-trivial. As illustrative examples we study density waves of two different sorts: CDW and SDW with ordering vector  $(\pi,\pi)$, for which the unit cell is doubled, i.e., there are two bands and the first Brillouin zone is halved.

In a ``site-centered'' CDW, the on-site energies are $+\Phi$ on the even and $-\Phi$ on the odd sublattice.  An SDW is similarly constructed, although now  $\Phi$ is a vector in spin space, whose direction defines the axis of quantization, such that the on-site energy is opposite for spin up and down electrons. The CDW and SDW share the same dispersion:
\ba
\epsilon_{\pm} (\textbf{k}) =&& -4t^{\prime} \cos k_{x} \cos k_{y} \\
&&\pm \sqrt{|\Phi|^{2}+4t^{2}[1+\cos(k_{x}\!+\!k_{y})][1+\cos(k_{x}\!-\!k_{y})]}\,,
\nonumber
\ea
where $\pm$ correspond to ``conduction'' and ``valence'' bands. For $|\Phi|<2|t^{\prime}|$, the two bands overlap in energy, and even for $n_{\textmd{el}}=1$, the system is a two-band metal. For $|\Phi|>2|t^{\prime}|$, however, the system is a band insulator when $n_{\textmd{el}}=1$.
The evolution of the Fermi surface as a function of $\Phi$ is shown in Fig.~\ref{dwfshalf}.  The density-wave order causes a reconnection of the Ferm surface, resulting for $n_{\textmd{el}}=1$ in two inequivalent hole pockets which enclose the (folded) zone edge centers at $(\pm\pi/2,\pm\pi/2)$, and an electron pocket, whose area is equal to the sum of those of the hole pockets, enclosing the zone corners at $(\pi,0)$ and $(0,\pi)$.  For somewhat smaller $n_{\textmd{el}}$,  the relative size of the electron pocket decreases such that, for large enough $\Phi$, only the hole pockets survive. For SDW, spin-rotational symmetry is broken. Hence, the three-fold degeneracy of the spin-1 pairs is lifted, so we must treat separately the $S_z=0$ pairing between opposite-spin electrons (obtained from the eigenstates of $\Gamma_{\uparrow\downarrow}$) and the doubly degenerate $S_z=\pm 1$ pairing between like-spin electrons (obtained from the eigenstates of $\Gamma_{\uparrow\uparrow}$).

\subsubsection{Half filling}

\begin{figure}[t]
\subfigure[]{
\includegraphics[scale=0.8]{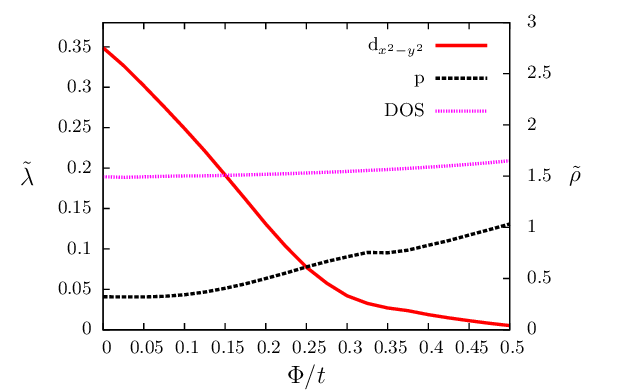}}\\
\subfigure[]{
\includegraphics[scale=0.8]{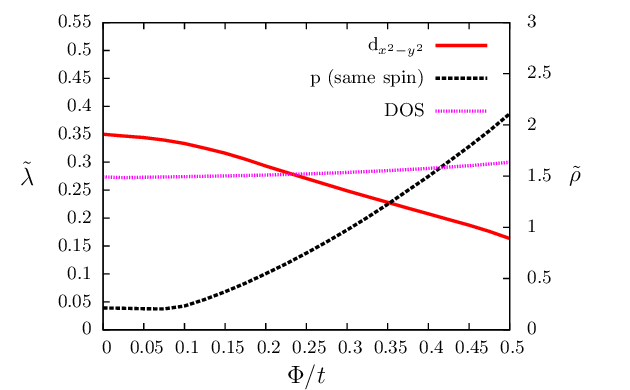}}
\caption{Pairing strengths and the density of states ($y$ axis to the right)  for $n_{\textmd{el}}=1$ and $t^{\prime}=-0.3t$ in the presence of the (a) $(\pi,\pi)$-CDW and (b) $(\pi,\pi)$-SDW.}
\label{dwpairinghalf}
\end{figure}

\begin{figure}[t]
\subfigure[]{
\includegraphics[scale=0.21]{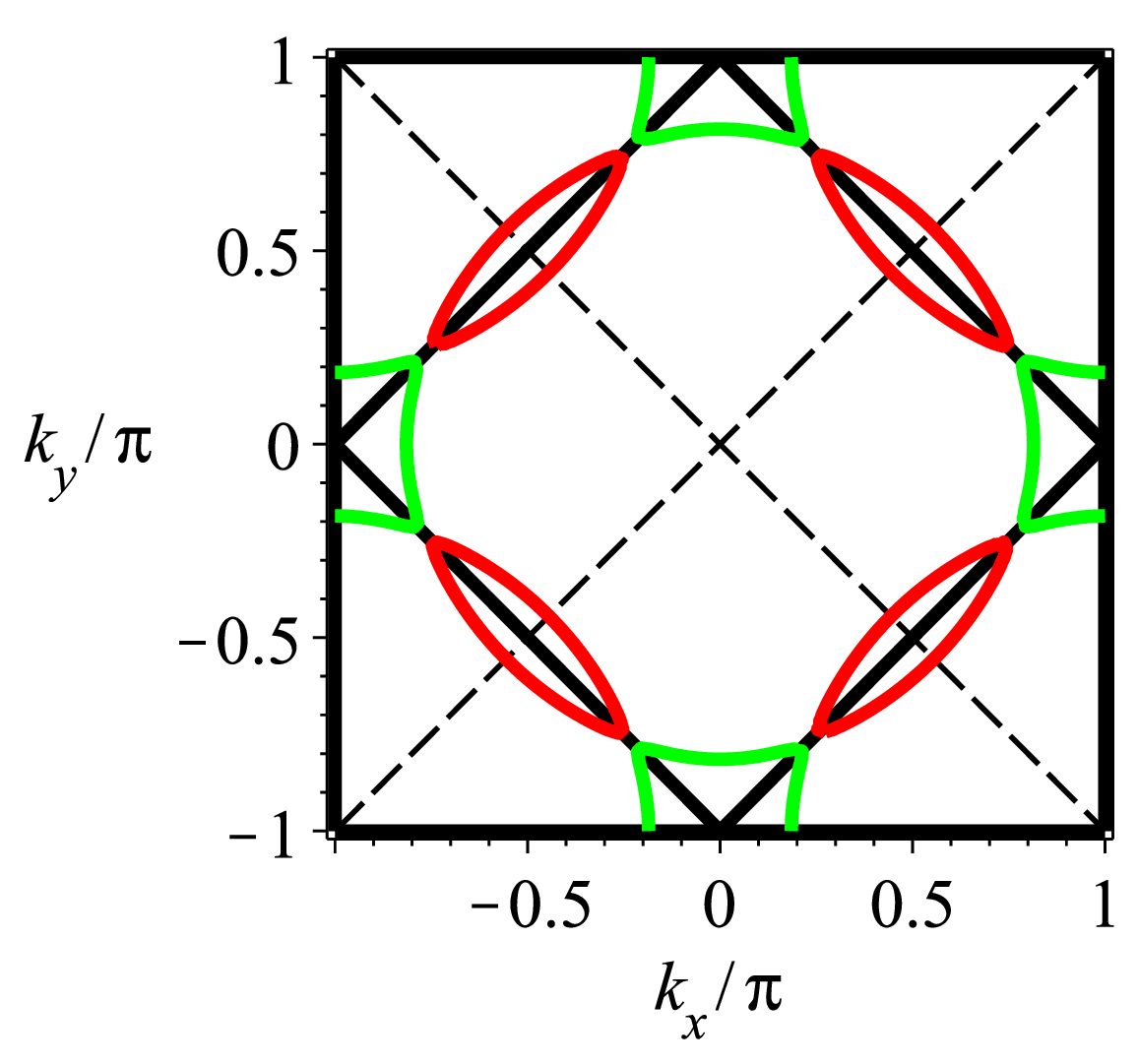}}
\subfigure[]{
\includegraphics[scale=0.21]{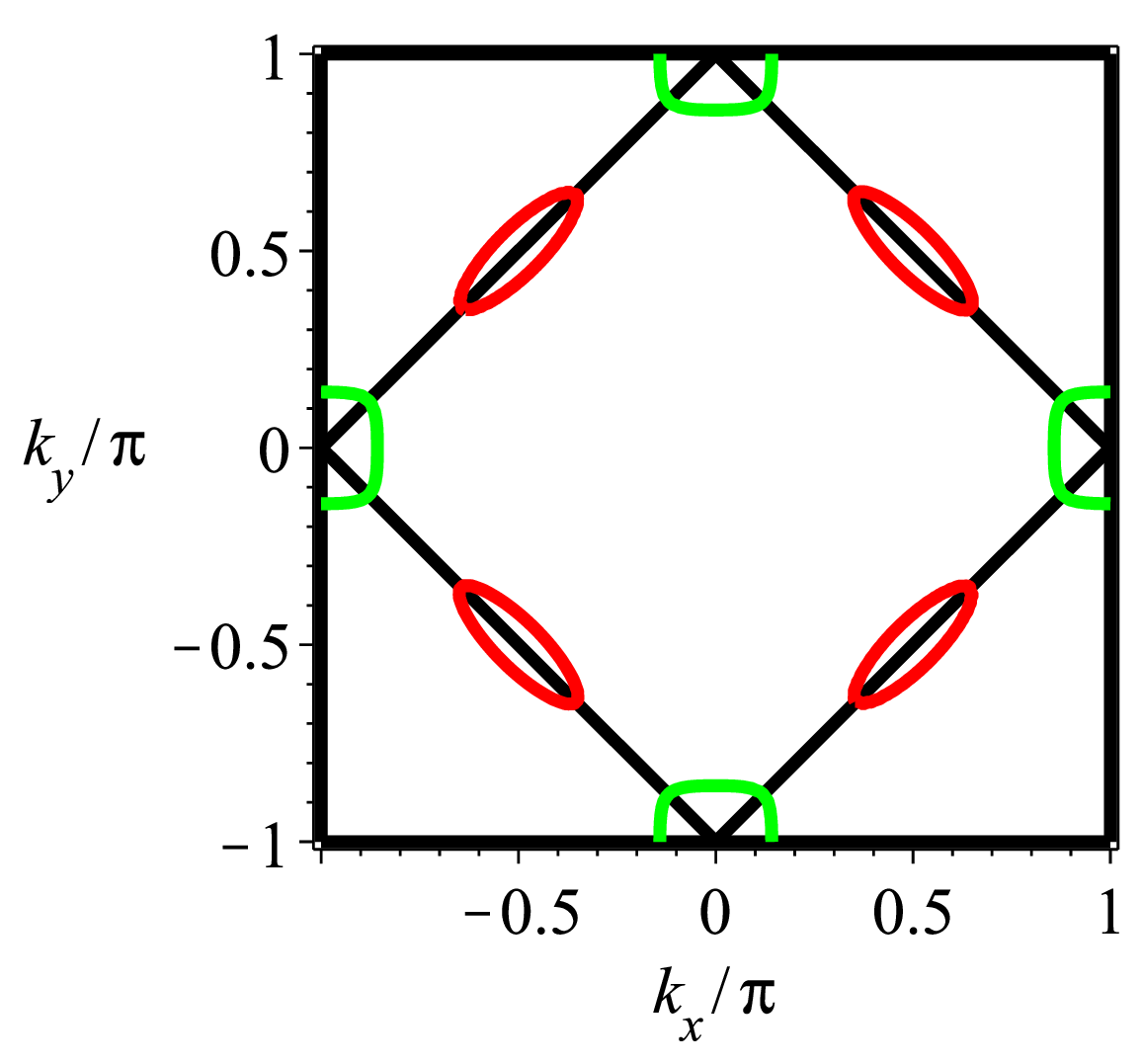}}
\caption{The Fermi surface in the presence of the $(\pi,\pi)$-CDW/SDW background for half filling and $t^{\prime}=-0.3t$. The inner black square rotated by $45^{\circ}$ represents the folded zone. (a) $\Phi=0.08$: Dashed lines denote the nodes of the $d_{x^{2}-y^{2}}$-wave gap. (CDW and SDW) (b) $\Phi=0.4$ ($p$ wave for CDW and SDW.).}
\label{dwfshalf}
\end{figure}

For $n_{\textmd{el}}=1$, for both kinds of density waves, the dominant pairing symmetry remains $d$-wave for a range of $\Phi$, but gives way to $p$-wave pairing for  $|\Phi|> \Phi_c\sim0.3t$ as shown in Fig. \ref{dwpairinghalf}. In the SDW case, the dominant $p$-wave solution involves pairing of like-spin electrons. Since $C_4$ rotational symmetry is preserved, the $p$-wave solutions correspond to a two-dimensional representation---the particular pattern of $p$-wave pairing below $T_c$ is determined by non-linear interactions not treated in the perturbative RG analysis, although it is likely that $p\pm ip$ pairing will maximize the condensation energy. For the CDW case, a basis can be found in which one component of the $p$-wave gap  lives predominantly on one of the hole pockets whose minor axis coincides with the associated $p$-wave nodal line, while the other component is associated with the other hole pocket.  For the SDW case, the $p$-wave pairs reside primarily on the electron pocket (see Fig.~\ref{dwfshalf}).

We also see that $d$-wave pairing is uniformly suppressed with increasing $\Phi$;  in this sense, the density-wave order and superconductivity ``compete.''  However, the $p$-wave pairing strength is an increasing function of $\Phi$, so in the regime of dominant $p$-wave pairing, density-wave order enhances superconductivity.
The evolution of the pairing strengths in Fig.~\ref{dwpairinghalf} is particularly notable:  at the right-hand edge of the figure, the size of the Fermi pockets is tending to zero upon approach to the metal-insulator transition which occurs at $\Phi=2|t^\prime|=0.6t$. Counterintuitively, the strength of the $p$-wave pairing grows all the way to the border of the insulating phase. A similar behavior has been found for the case of the pnictide LiFeAs, where, taking the two-dimensional ($k_z=0$) limit of the electronic model, a very small pocket triggers a considerable propensity to ferromagnetic fluctuations and hence $p$-wave superconductivity. \cite{platt-11prb235121,brydon-11prb060501}

This peculiarity derives from the property of two dimensions that even when the Fermi surface is arbitrarily small around a quadratic band edge, the density of states is finite and hence there are enough initial and final states available for scattering. This alone is not sufficient to explain the stability of a nodal ($p$-wave) superconducting solution in this limit. Nonzero pairing strength with one or more sign change within an infinitesimally small pocket means that the effective interaction must be substantially ${\bf k}$-dependent over this small range. This kind of singularity can arise due to particle-hole pairs excited around a similarly small pocket so that the Fermi function in Eqs. (\ref{effintupdown}) or (\ref{effintupup}) changes sensitively as the tiny momentum transfer varies. (When a Fermi pocket is centered at a time-reversal invariant momentum, which is indeed true for our case, not only $\hat{\textbf k}-\hat{\textbf k}^{\prime}$ but also $\hat{\textbf k}+\hat{\textbf k}^{\prime}$ modulo the periodicity of the folded BZ is of the order of the pocket size.)

We elaborate further on how the singular momentum structure described above can arise. Suppose we focus on a term in Eqs. (\ref{effintupdown}) or (\ref{effintupup}) corresponding to a case where the intermediate states arise from an infinitesimally small pocket. Then, the integrand is non-vanishing only in a tiny region where the difference of Fermi functions is nonzero, and the product of two bare vertex functions is essentially constant over this region. The remaining integral is then simply proportional to the usual particle-hole susceptibility for a quadratic dispersion, which is given by the following analytic expression:
\begin{equation}
\label{chiquadratic}
\chi(\textbf{q}) = -\rho_{0}\left(1-\textmd{Re}\sqrt{1-\frac{1}{\alpha(\textbf{q})}}\right)\,,
\end{equation}
where $\alpha(\textbf{p})\equiv(q_{x}/2k_{F,x})^{2}+(q_{y}/2k_{F,y})^{2}$ and $\textbf{q}=\hat{\textbf k}\pm\hat{\textbf k}^{\prime}$ is the momentum transfer. $2k_{F,x}$ and $2k_{F,y}$ are major and minor axes of an elliptical Fermi pocket. $\rho_{0}$ is the density of states per spin. The above expression is constant for $\alpha(\textbf{q})<1$ and otherwise monotonically decreases as a function of $\alpha(\textbf{q})$.

For a non-trivial momentum structure to be possible at all, the pocket responsible for the structure of $\chi$ must be ``smaller'' than the pocket on which the pairs reside---or, more precisely, the pocket which mediates the effective interaction must not be able to enclose the pocket on which the pairing takes place if the two were put on top of one another. (Among other things, this means that a single pocket cannot both be the home of the pairing electrons and of the particle-hole excitations that mediate the pairing interactions.) Conversely, in order for the pairing strength to be substantial, the pocket on which the pairs reside must not be too much larger than the one which mediates the interaction \cite{raghu-11prb094518} since if it were too large, the induced interaction would be small for all but the smallest momentum transfer pair-scattering processes---i.e., the effective interaction would be very weak.

In our example of density waves, there is an additional simplification. At the folded BZ boundary, where the tiny Fermi pockets are located, the Bloch states from each band reside strictly in a single sublattice. [This can be inferred from the fact that the band energy does not depend on $t$ wherever $k_x =\pm(\pi\pm k_y)$.]
This, along with the property of the Hubbard interaction that it is diagonal in the sublattice index, render Fermi pockets living in different sublattices decoupled from each other. The consequence is that in the CDW state, the bare interaction between the electron  and  hole pockets does not exist, whereas in the SDW state, this is the only allowed bare interaction.

To second order in $U$ in the CDW case, the effective intrapocket interaction acquires singular momentum structure only when the electron pairing occurs on one of the hole pockets, and the virtual particle-hole pairs are from the other hole pocket. In the SDW,  the interaction of like spins within the electron pocket mediated by virtual pairs from one of the hole pockets and the converse process produce $\bf{q}$-dependent effective interactions.  However, the former are generally more effective. As can be seen from Fig.~\ref{dwfshalf}b, each hole pocket almost fits into the electron pocket, and therefore, does a poor job in generating momentum-dependent interactions at the electron pocket. This observation leads us to expect that the nodal $p$-wave solution should predominantly reside in the hole pockets for CDW and the electron pocket for like-spin pairing in the SDW, which is in perfect agreement with what we find numerically.

\subsubsection{$n_{\textmd{el}}=0.8$}

\begin{figure}[t]
\subfigure[]{
\includegraphics[scale=0.8]{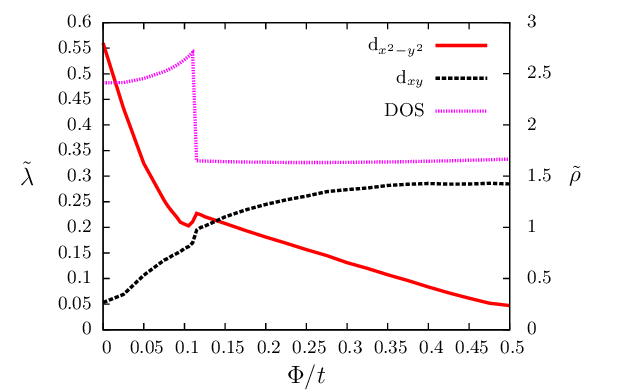}}\\
\subfigure[]{
\includegraphics[scale=0.8]{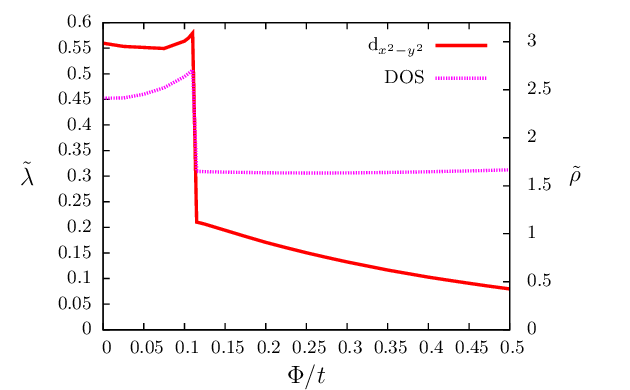}}
\caption{
Pairing strengths and the density of states (y-axis to the right) for $n_{\textmd{el}}=0.8$ and $t^{\prime}=-0.3t$ in the presence of the (a) $(\pi,\pi)$-CDW and (b) $(\pi,\pi)$-SDW.
}
\label{dwpairing08}
\end{figure}

\begin{figure}[t]
\subfigure[]{
\includegraphics[scale=0.21]{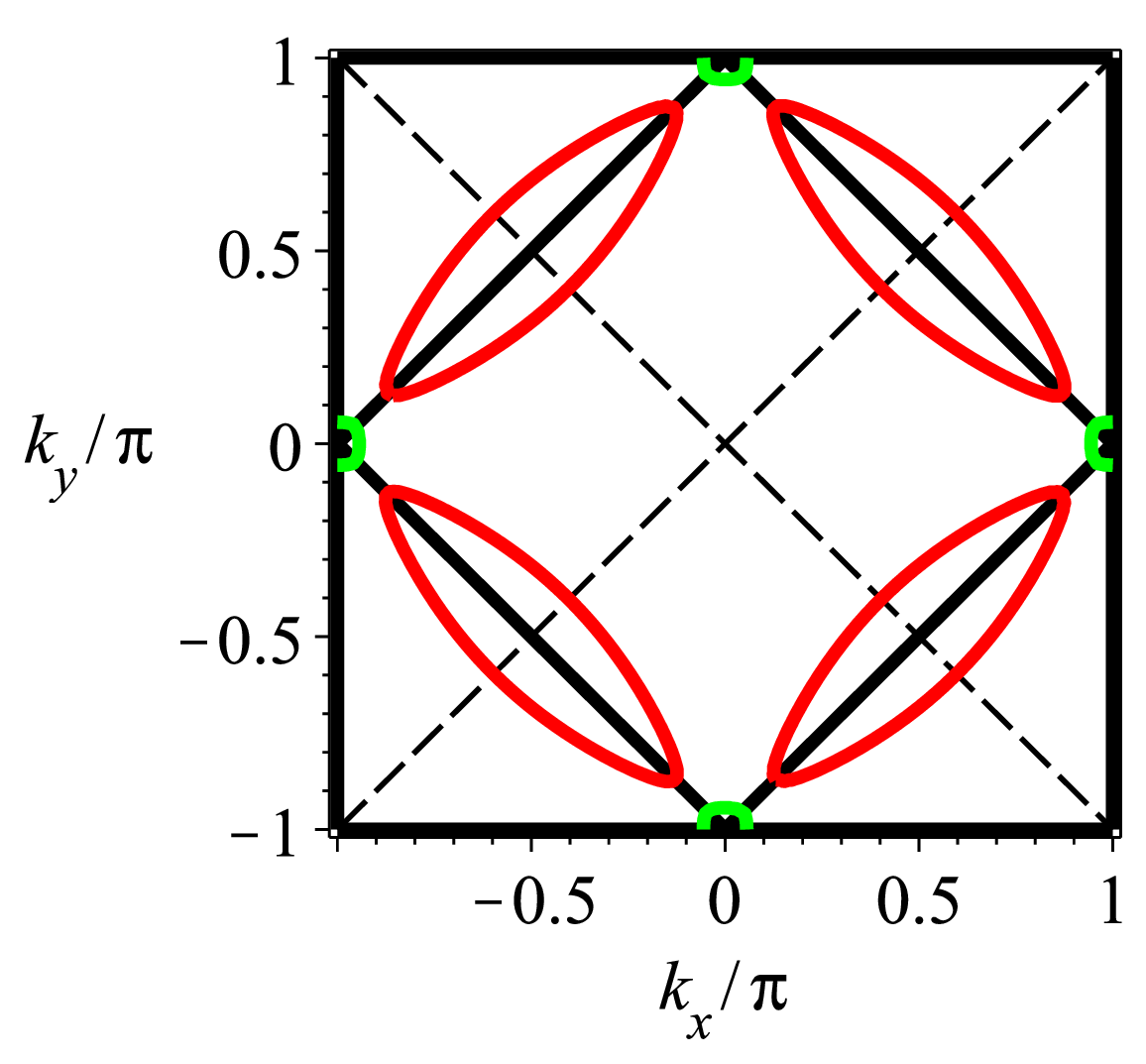}}
\subfigure[]{
\includegraphics[scale=0.21]{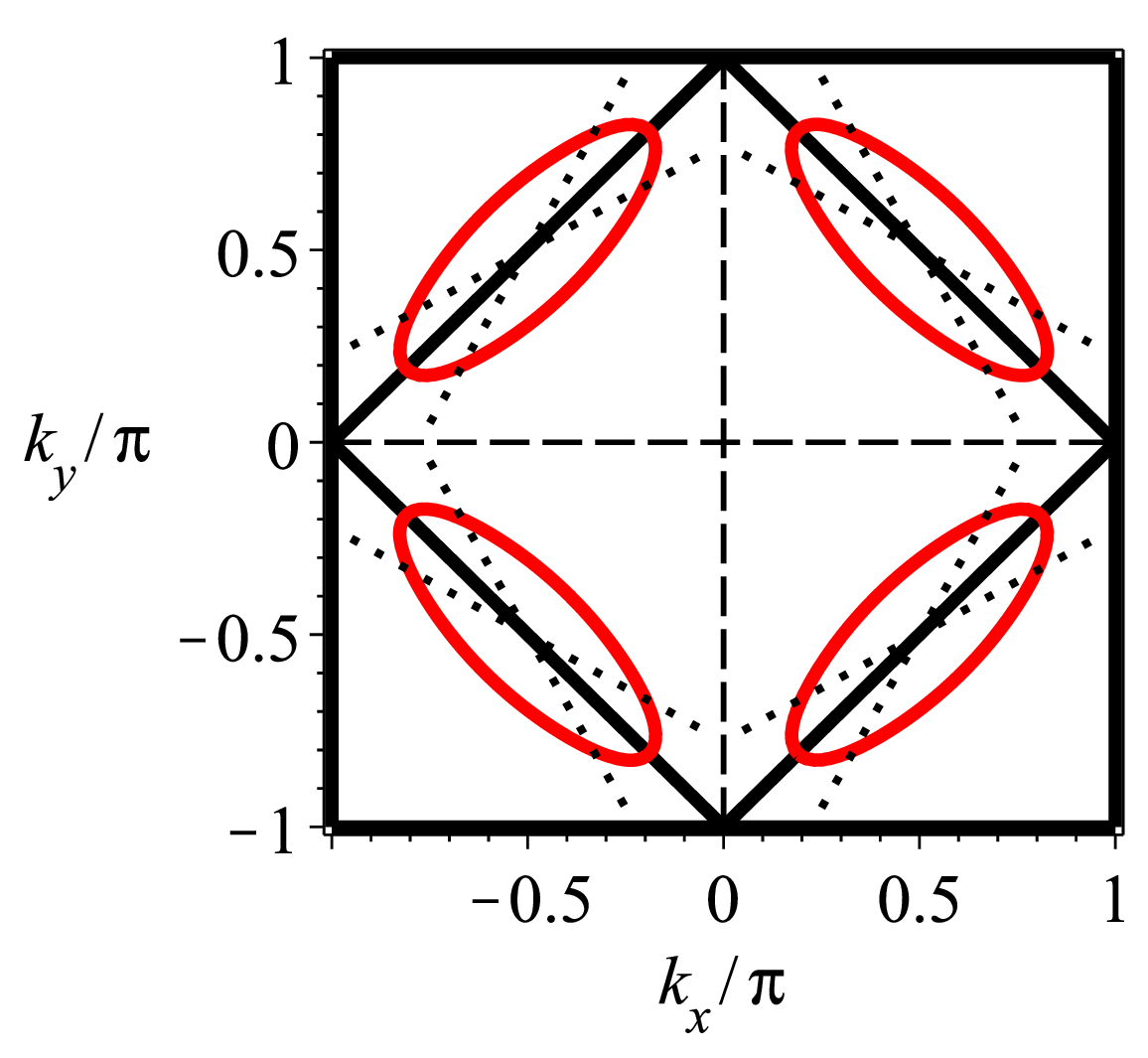}}
\caption{The Fermi surface in the presence of the $(\pi,\pi)$-CDW/SDW background for $n_{\textmd{el}}=0.8$ and $t^{\prime}=-0.3t$. The inner black square rotated by $45^{\circ}$ represents the folded zone. (a) $\Phi=0.08$: Dashed lines denote the nodes of the $d_{x^{2}-y^{2}}$-wave gap. (CDW \& SDW) (b) $\Phi=0.4$: Dashed and dotted lines denote the symmetry-related ($d_{xy}$) and accidental nodes, respectively, for CDW. ($d_{x^{2}-y^{2}}$ for SDW.)}
\label{dwfs08}
\end{figure}

It is noteworthy that the hole-doped system($n_{\textmd{el}}=0.8$) shows a qualitatively different behavior as $\Phi$ increases compared to the case of half filling. For the CDW, the pairing symmetry is $d_{xy}$ for $\Phi\gtrsim0.14$, as can be seen from Fig.~\ref{dwpairing08}. We have found that the $d_{xy}$ pairing solution predominantly lives in the hole pockets, and accidental nodes form within each pocket (see Fig.~\ref{dwfs08}). The singularity seen in Fig.~\ref{dwpairing08}(a) around $\Phi=\Phi_{c}\approx0.11$ marks the point where the electron pocket vanishes. For SDW, there is no transition in the pairing symmetry, but superconductivity is significantly suppressed when the electron pocket disappears.

We see that both kinds of density-wave orders  eliminate some (or for large enough, $\Phi$, all) of the Fermi surface in the ``antinodal'' region, much the way these states are eliminated by the psuedo-gap in underdoped cuprates.  Moreover, just as the pseudo-gap seems to suppress $T_c$ in the cuprates, the presence of density-wave order in our calculations tends strongly to suppress $d_{x^{2}-y^{2}}$ superconductivity. However, in the cuprates, there is certainly no sign of the change of the symmetry of the superconducting state which we find when the CDW order is strong enough to fully gap the antinodal portion of the Fermi surface.  Despite the emerging evidence of CDW order in at least some portions of the pseudo-gap regime, this presents a significant barrier to any theory that identifies the pseudo-gap simply with a CDW gap.

Notice that there is a striking contrast between the CDW and SDW cases, in how the pairing strength of the $d_{x^{2}-y^{2}}$-wave behaves at $\Phi$=$\Phi_{c}$, at which the electron pocket vanishes. From Fig.~\ref{dwpairing08}, we see that there is only a continuous singularity for the CDW case, while there is a sharp discontinuity in SDW case. The reason for this difference is that in the former, the tiny electron pocket (just before it vanishes) can affect superconductivity only by providing virtual electron-hole pairs mediating the effective interaction, whereas in the latter, a finite fraction of the pairing solution actually resides on the vanishingly small pocket.

\subsection{Checkerboard model}
\label{checkerboard}

\begin{figure}[t]
\includegraphics[scale=0.8]{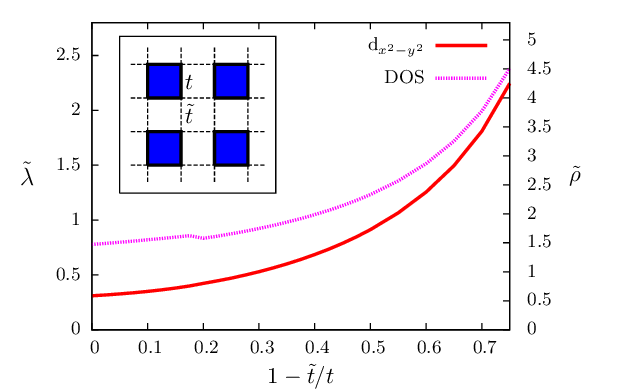}
\caption{Pairing strengths and the density of states ($y$ axis to the right) when the checkerboard pattern (as described in the inset) is imposed for $n_{\textmd{el}}=0.8$. $\tilde t$ denotes the weak hopping amplitude between different plaquettes.
}
\label{chkbdpairing}
\end{figure}

The checkerboard model  is defined by a form of $2\times 2$ plaquette-density-wave order on the square lattice, as shown in the inset of Fig.~\ref{chkbdpairing}. Specifically, the model we consider is obtained from the  square lattice model  by weakening  half the nearest-neighbor bonds ($t \rightarrow \tilde t<t$) in such a way that each quadrupled unit cell contains a square plaquette formed by strong bonds, with different plaquettes connected by weak bonds. For simplicity, next-nearest-neighbor hoppings are set to zero ($t'=0$). The dispersion of the resulting four bands is given by
\ba
\epsilon_{r,r^\prime}(\textbf{k})&=-\Big[r\sqrt{t^{2}+\tilde t^{2}+2t\tilde t\cos 2k_{x}} \\
\nonumber
&+r^\prime\sqrt{t^{2}+\tilde t^{2}+2t\tilde t\cos 2k_{y}}\Big].
\ea
where $r=\pm$ and $r^\prime=\pm$.

\begin{figure}[b]
\subfigure[]{
\includegraphics[scale=0.21]{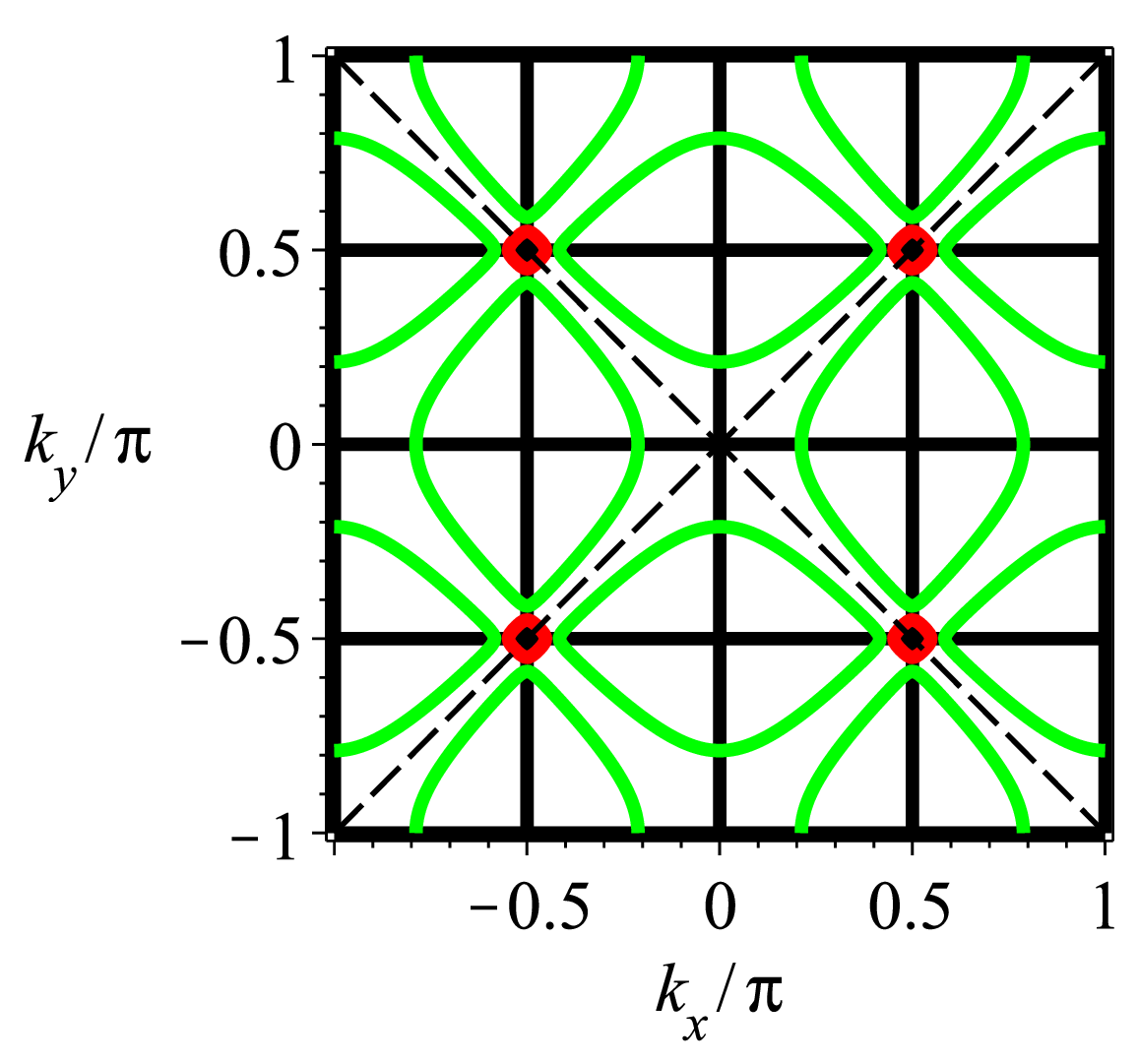}}
\subfigure[]{
\includegraphics[scale=0.21]{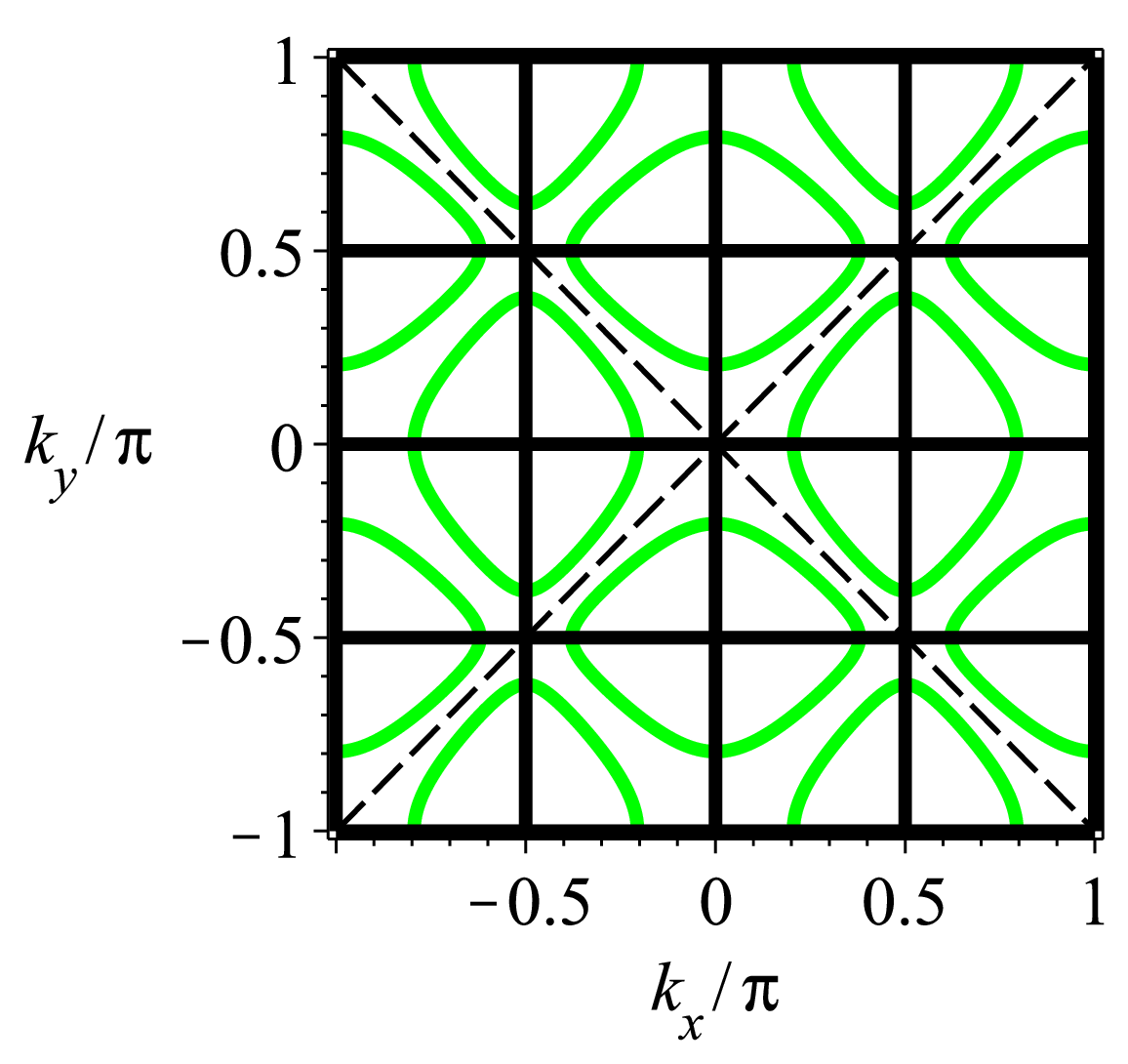}}
\caption{The Fermi surface in the presence of the checkerboard pattern for $n_{\textmd{el}}=0.8$. (a) $\tilde t=0.9t$. (b) $\tilde t=0.6t$. The hole pocket is present only in the former case. Dashed lines represent the nodes of the $d_{x^{2}-y^{2}}$-wave gap.}
\label{chkbdFS}
\end{figure}

Because we have not included any second-neighbor hopping, the system is particle-hole symmetric at half-filling and thus has a number of non-generic features of its band structure; the density of states is divergent due to a Van Hove singularity, and the Fermi surface is perfectly nested with nesting vector $(\pi,\pi)$, just as the square lattice with  uniform nearest-neighbor hopping amplitudes. However, for $n_{\textmd{el}} <1$, the properties of the system are more robust.  As shown in Fig. \ref{chkbdFS},  there are two symmetry-related electron pockets originating from the $(+,-)$ and the $(-,+)$ bands that enclose, respectively, the points $(0,\pi/2)$ and $(\pi/2,0)$. One hole pocket from the $(+,+)$ band exists, enclosing $(\pi/2,\pi/2)$, for a sufficiently weak checkerboard pattern, i.e., for $\tilde t/t$ close enough to 1,  but this pocket is lost for  $\tilde t/t$ smaller than a critical ratio, $(\tilde t/t)_c$.

Figure \ref{chkbdpairing} shows the pairing strength for $n_{\textmd{el}}=0.8$ as $\tilde t/t$ is varied. $d_{x^{2}-y^{2}}$ is always the dominant pairing channel, while the strength of pairing grows as $\tilde t$ is decreased from $t$. This pairing symmetry is naturally expected. There is a large density of states at the electron pockets as they lie close to the Van Hove singularity at $(0,0)$. The $d_{x^{2}-y^{2}}$ permits a gap structure that for $\tilde{t}/t<(\tilde{t}/t)_{c} \approx 0.81$ does not have any  nodes which intersect a Fermi surface,  but none-the-less has the favored sign changes between different electron pockets.

We see that the pairing strength as a function of $\tilde{t}$ is smooth at $(\tilde{t}/t)_c$, below which the hole pocket does not exist. This reflects the fact that the hole pocket, especially when it is vanishingly small, participates negligibly to superconductivity. Formally, the loss of the contribution of intermediate electron-hole pairs from this pocket causes a weak singularity in the pairing strength. However, it is not surprising that this singularity is not visible in this particular case because the density of states of the hole pocket is small, as can be seen from Fig.~\ref{chkbdpairing} by the small size of the jump in DOS at $(\tilde{t}/t)_c$.

In the context of an attempt to determine whether there is an ``optimal inhomogeneity for superconductivity,'' this model has been studied previously for intermediate values of $U$ using exact diagonalization, \cite{yao-07prb161104r} DMRG, \cite{karakonstantakis-11prb054508} CORE, \cite{baruch-10prb134537} and DCA. \cite{tremblay-11prb054545,maier-08prb020504} Similar conclusions have been reached concerning an enhancement of $T_c$ for intermediate strength of the checkerboard potential in all but the DCA results.  (We speculate that the DCA results are probably artifacts of the small cluster sizes used in those calculations.)
In the weak-coupling regime, the rise of $T_c$ with decreasing $\tilde{t}/t$ is pronounced, consistent with all the studies other than the DCA.  However, the enhancement of superconductivity we find is largely in agreement with what a simple reduction of bandwidth would result in, i.e., $\rho \propto 1/W$ and $V_{\textmd{eff}}\propto U^{2}/W$, and therefore, $\lambda = \rho V_{\textmd{eff}}\propto U^{2}/W^{2} \propto \rho^{2}$.

\subsection{Bilayer model}

\begin{figure}[t]
\includegraphics[scale=0.8]{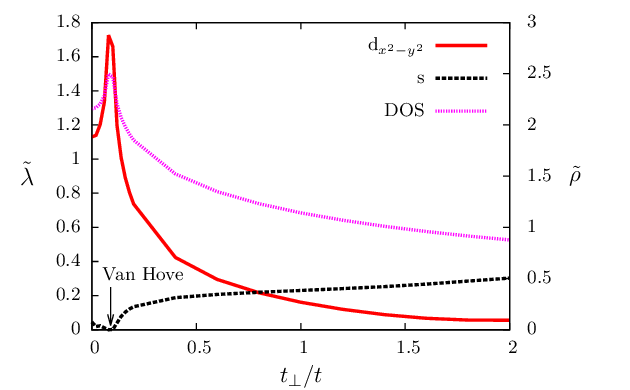}
\caption{Pairing strengths and the density of states ($y$ axis to the right) in the bilayer model for $n_{\textmd{el}}=0.95$ and with no second-neighbor hopping.}
\label{bilayerpairing}
\end{figure}

The dispersion for the bilayer square lattice with (intra-layer) nearest-neighbor hopping $t$ and interlayer hopping $t_{\perp}$, is given as follows:

\begin{equation}
\epsilon_{\pm}(\textbf{k}) = -\Big[2t(\cos k_{x} + \cos k_{y}) \pm t_{\perp}\Big]\,.
\end{equation}

\begin{figure}[b]
\subfigure[]{
\includegraphics[scale=0.21]{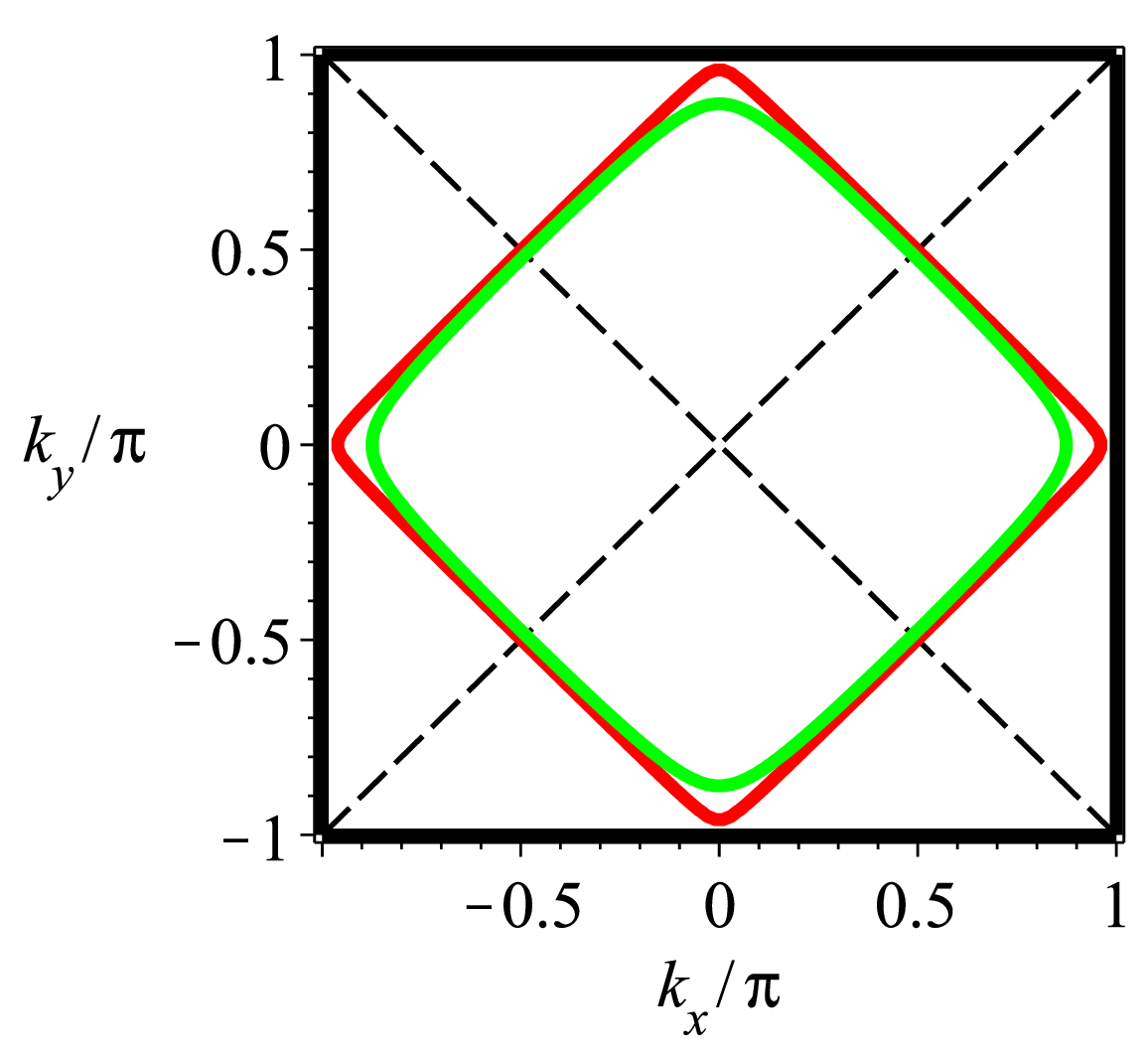}}
\subfigure[]{
\includegraphics[scale=0.21]{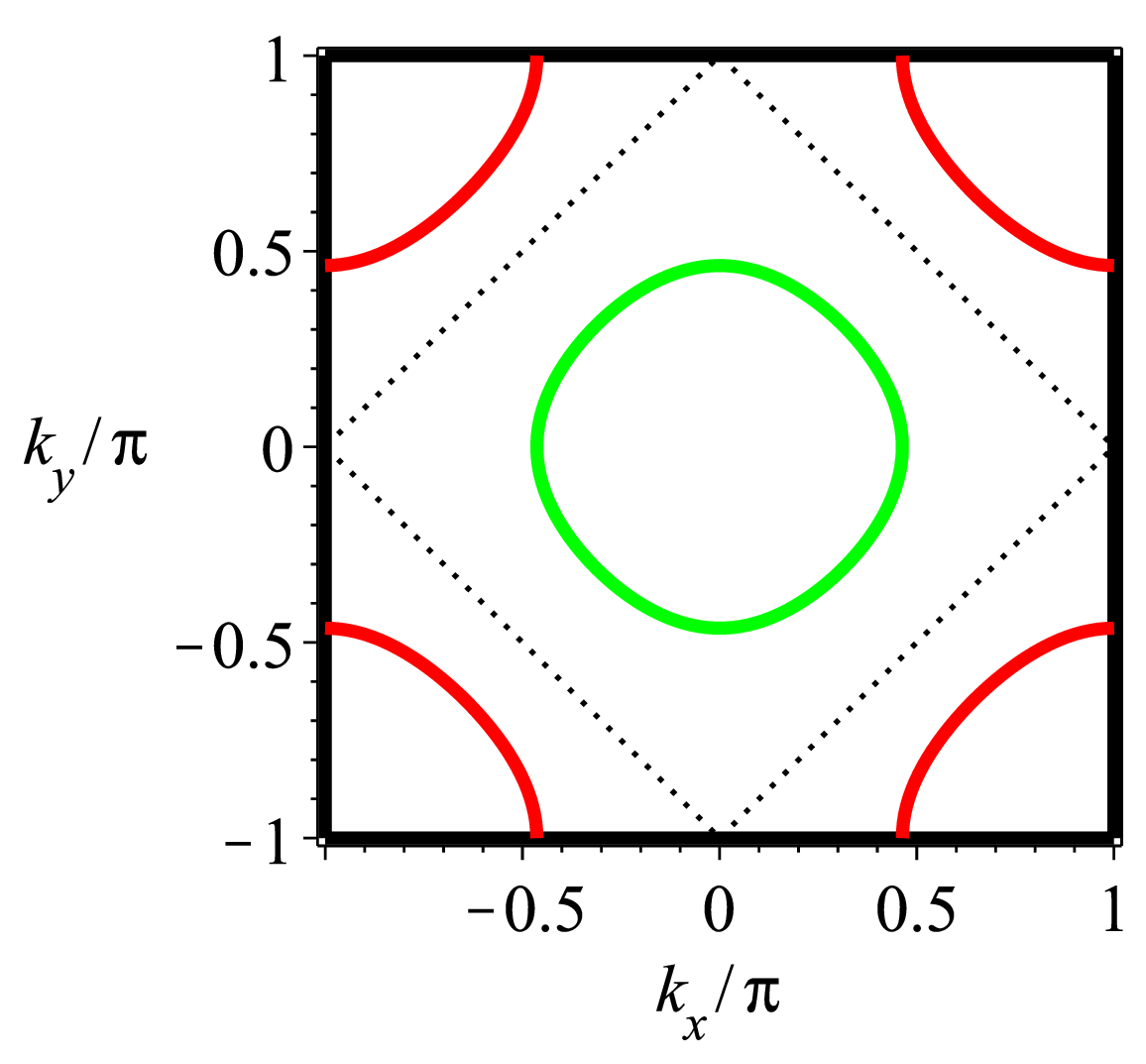}}
\caption{The Fermi surface for the Bilayer model for $n_{\textmd{el}}=0.95$. (a) $t_{\perp}=0.07t$: Dashed lines denote the nodes of the $d_{x^{2}-y^{2}}$-wave gap. (b) $t_{\perp}=2.0t$: The $s_{\pm}$-wave gap changes its sign across the dotted lines.}
\label{bilayerFS}
\end{figure}

The $(+)$ and $(-)$ band correspond to the bonding and anti-bonding states between the layers, respectively. As in the case of the checkerboard pattern, we again have a particle-hole symmetric system with  perfect nesting for $(\pi,\pi)$ at half filling, but the Van Hove singularity does not coexist with the perfect nesting unless $t_{\perp}=0$. We will consider the case in which $n_{\textmd{el}}$ is slightly less than 1, where this nesting is imperfect, but none-the-less results in a significant peak in the effective interaction at $(\pi,\pi)$.

When $t_{\perp}\approx 0$, the bilayer essentially behaves as two independent copies of the square lattice. There are two almost identical electron-like Fermi surfaces from each of the bands, where the one from the $(+)$ band is slightly larger. On the other hand, for sufficiently large $t_{\perp}$, the Fermi surface from the $(+)$ band closes around  $(\pi,\pi)$ and becomes a hole pocket, while the one from the $(-)$ band remains an electron pocket around $(0,0)$ (see Fig.~\ref{bilayerFS}). A Van Hove singularity occurs at a critical value of $t_\perp/t$ at the border between  these two regimes.

Figure \ref{bilayerpairing} shows the pairing strengths of the bilayer system as a function of $t_{\perp}/t$ for $n_{\textmd{el}}=0.95$. The pairing symmetry is $d_{x^{2}-y^{2}}$ for $t_{\perp}/t\approx0$, which is inherited from the case of $t_{\perp}=0$. Superconductivity is enhanced around the Van Hove singularity at $t_{\perp}/t \approx 0.09$, and there is a transition around $t_{\perp}/t \sim 1$, past which the dominant form of superconductivity is an unconventional $s$-wave that changes sign between the electron and the hole pocket. ($s_{\pm}$) This pairing solution makes the best use of the large effective interaction at the momentum transfer of $(\pi,\pi)$ due to the approximate nesting. Note that the approximate degeneracy of $s_\pm$ and $d$-wave at the instability level does not necessarily imply the existence of an intermediate $s+id$ phase. \cite{zhang-09prl217002,tesanovic-10prb134522,thomale-11prl117001,platt-12prb180502,hirschfeld-11rpp124508}

\section{The mechanism of unconventional superconductivity}
\label{discussion}

Models with weak, short-range electronic repulsions admit to a controlled solution, but  are somewhat artificial;  in real materials, interactions are always complicated and of substantial magnitude.  In this final section, we extrapolate the insights we have obtained from this controlled limit to venture some more general ``principles of unconventional superconductivity,'' and to make inferences concerning the physics of a variety of unconventional superconductors.

While conventional superconductors differ in many salient details, enough essential features are shared that it makes sense to talk about a single ``conventional mechanism of superconductivity;''  there is an  induced attraction between electrons due to the exchange of phonons that is highly retarded, and so able to overcome the generally stronger but instantaneous bare repulsion between electrons.  Because the electron-phonon coupling is typically relatively local in space, the resulting superconducting gap function is weakly structured in ${\bf k}$-space, although it is strongly frequency dependent.  Because $T_c$ is small compared to the phonon-energy, which is in turn small compared to the Fermi energy, pairing and phase coherence occur essentially simultaneously, and mean-field theory is thus extremely accurate.

The essential feature of the unconventional mechanism explored here is that the bare repulsion is short-ranged while the induced attraction
is longer-ranged.  The resulting gap-function must be strongly ${\bf k}$-dependent, such that  the average   over the Fermi surface of the gap is small compared to the root-mean-square gap, $\big|\overline{\Delta({\bf k})}\big|^2\ll\overline{|\Delta({\bf k})|^2}$.  However, the symmetry of the gap function is {\it not} essential;  depending on details of the band structure, the same ``mechanism'' can give rise to various forms of sign-changing $s$-wave superconductivity, not to mention both triplet and singlet pairing.  Indeed,  at least in the strongly nematic case, we have seen that a singlet and triplet state can be nearly degenerate over a broad range of parameters, driven by precisely the same interactions.

There has been considerable focus on unconventional pairing produced by fluctuations associated with a nearby ordered state, especially with a spin-density-wave state of one sort or another.  In the weak-coupling limit, except for exceptionally fine-tuned band structures, the correlations associated with any putative density-wave states are always weak.  Strong spin fluctuations (or CDW, nematic, orbital current, or dDW fluctuations)  certainly occur under broad circumstances as the interactions get stronger, but the absence of a small parameter makes well-controlled theory impossible.  However, it is physically plausible that,  as the susceptibility towards some particular SDW (or some other ordered) state grows stronger, the induced attractions between electrons are enhanced  and with them the pairing scale.

Various appealing approximation schemes have been used to explore the pairing in the intermediate-coupling regime, of which fRG \cite{halboth-00prb7364,honerkamp-01prb035109,wang-09prl047005,thomale-11prl187003,metzner-12rmp299,hur20091452} or parquet RG \cite{hur20091452,tesanovic-09prb024512,chubukov-08prb134512,chubukov2009640,chubukov13} are the most directly related to the methods used in the present paper.  These approaches  generally give results similar to the weak-coupling approach in the appropriate limit.  Moreover, as a function of the strength of the couplings, these approaches   usually lead to a gap function which
does not change its symmetry or nodal structure as the coupling strength is increased.   Descriptive words based on such intermediate-coupling approaches, such as ``spin-fluctuation exchange mechanism,'' \cite{scalapino12rmp1383}  may be applicable in some specific cases, although the spin fluctuations in question are typically rather short-range correlated.  More importantly,  in our opinion, this nomenclature obscures a more basic commonality, in which the short-range repulsive interactions between electrons is the dominant feature.

\begin{acknowledgments}

We thank A.~V.~Chubukov and D. J. Scalapino for discussions and comments. RT acknowledges support by SPP-1458.  This work was supported
in part by the Office of Basic Energy Sciences,
Materials Sciences and Engineering Division of the U.S.
Department of Energy under Contract No.  AC02-76SF00515 (WC, SAK, SR),
and the Alfred P. Sloan Foundation (SR).
\end{acknowledgments}

\bibliography{Hubbard_Band_Jul8}

\end{document}